\documentclass[12pt,preprint]{aastex61}
\usepackage{natbib}
\usepackage{hyperref} 
\usepackage{amsmath}
\newcommand{\um}{$\mu$m}

\newcommand{\Msun}{M$_{\odot}$}

\newcommand{\kms}{km~s$^{-1}$}

\newcommand{\cmc}{${\rm cm}^{-3}$}

\newcommand{\hii}{\mbox{$\mathrm{H\,{\scriptstyle {II}}}$}}

\newcommand{\nthpnt}{N$_{2}$H$^{+}$}
\newcommand{\nthp}{N$_{2}$H$^{+}$\,(1--0)}
\newcommand{\hcopnt}{HCO$^{+}$}
\newcommand{\hcop}{HCO$^{+}$\,(1--0)}

\newcommand{\htcopnt}{H$^{13}$CO$^{+}$}
\newcommand{\htcop}{H$^{13}$CO$^{+}$\,(1--0)}

\newcommand{\hcn}{HCN (1--0)}

\shorttitle{MALT90 Collapse Signatures}
\shortauthors{Jackson et al.}

\begin{document}

\title{Asymmetric Line Profiles in Dense Molecular Clumps Observed in MALT90: Evidence for Global Collapse}

\author{James M. Jackson},
\affiliation{SOFIA Science Center, USRA, NASA Ames Research Center, Moffett Field CA 94045, USA}
\affiliation{School of Mathematical and Physical Sciences, University of Newcastle, University Drive, Callaghan NSW 2308, Australia}
\affiliation{Institute for Astrophysical Research, Boston University, 725 Commonwealth Ave.,  Boston MA 02215, USA}

\author{J. Scott Whitaker}
\affiliation{Physics Department, Boston University, 590 Commonwealth Ave., Boston MA 02215, USA; scott@bu.edu}

\author{J. M. Rathborne}
\affiliation{CSIRO Astronomy and Space Science,  P.O. Box 76, Epping NSW, 1710, Australia}

\author{J. B. Foster} 
\affiliation{Department of Astronomy, Yale University, P.O. Box 28101 New Haven CT 06520-8101, USA}

\author{Y. Contreras}
\affiliation{Leiden Observatory, Leiden University, PO Box 9513, 2300 RA Leiden, The Netherlands}

\author{Patricio Sanhueza}
\affiliation{Institute for Astrophysical Research, Boston University, 725 Commonwealth Ave., Boston MA 02215, USA}
\affiliation{National Astronomical Observatory of Japan, National Institute of Natural Sciences, 2-21-1 Osawa, Mitaka, Tokyo 181-8588, Japan}

\author{Ian W. Stephens} 
\affiliation{Institute for Astrophysical Research, Boston University, 725 Commonwealth Ave., Boston MA 02215, USA}
\affiliation{Harvard-Smithsonian Center for Astrophysics, 60 Garden St., Cambridge MA 02138, USA}

\author{S. N. Longmore}
\affiliation{Astrophysics Research Institute, Liverpool John Moores University, Egerton Wharf, Birkenhead CH41 1LD, UK}

\author{David Allingham}
\affiliation{School of Mathematical and Physical Sciences, University of Newcastle, University Drive, Callaghan NSW 2308, Australia}

\begin{abstract}
Using molecular line data from the Millimetre Astronomy Legacy Team 90 GHz Survey (MALT90), we
have searched the optically thick \hcop\, line for the ``blue asymmetry'' spectroscopic signature of infall motion in a large sample of high-mass, dense molecular clumps observed to be at different evolutionary stages of star cluster formation according to their mid-infrared appearance.  To quantify the degree of the line asymmetry, we measure the asymmetry parameter $A = {{I_{blue}-I_{red}}\over{I_{blue}+I_{red}}}$, the fraction of the integrated intensity that lies to the blueshifted side of the systemic velocity determined from the optically thin tracer \nthp.  For a sample of 1,093 sources,  both the mean and median of $A$ are positive ($A = 0.083\pm0.010$ and $0.065\pm0.009$, respectively) with high statistical significance, and a majority of sources (a fraction of $0.607 \pm 0.015$ of the sample) show positive values of A, indicating a preponderance of blue-asymmetric profiles over red-asymmetric profiles.   Two other measures, the local slope of the line at the systemic velocity and the $\delta v$ parameter of \citet{Mardones1997}, also show an overall  blue asymmetry for the sample, but with smaller statistical significance.  This blue asymmetry indicates that these high-mass clumps are predominantly undergoing gravitational collapse.  The blue asymmetry is larger ($A \sim 0.12$)  for the earliest evolutionary stages (quiescent, protostellar and compact H II region) than for the later H II region ($A \sim 0.06$) and PDR ($A \sim 0$) classifications. 

 \end{abstract}


\section{Introduction}

 High-mass stars  predominantly form in star clusters, which themselves originate in dense molecular clumps. The formation of high-mass stars is thus inextricably linked to the formation of star clusters and the evolution of dense molecular clumps.  The conversion of turbulent, self-gravitating, dense molecular gas clumps into high-mass stars
and star clusters remains an important open question in astrophysics (see reviews by \citealt{ZinneckerYorke2007}, \citealt{McKeeOstriker2007}, and \citealt{Motte2017}). 

Two competing theories, ``competitive accretion'' and ``turbulent core accretion,''  pose two distinct scenarios for high-mass
star-formation \citep{ZinneckerYorke2007}.
``Competitive accretion'' (cf., \citealt{Bonnell1997})  suggests that cores destined to become individual high-mass stars or binary pairs originally begin as  low-mass cores ($M_{core} \sim M_{Jeans} \sim 1$ \Msun\, for typical clump densities $n\sim10^4$ cm$^{-3}$ and temperatures $T\sim10$ K) that form via Jeans
fragmentation within a dense clump. Those cores which happen to lie near the center of a
collapsing dense clump receive a fresh supply of gas from afar as gas is funneled down
the clump's gravitational potential well, and these central cores then grow via accretion until they
eventually acquire enough mass to form high-mass stars. Alternatively, ``turbulent core accretion'' (cf., \citealt{KrumholzMcKeeKlein2005})
posits that turbulent eddies within a young clump
form virialized cores very early on, and that the highest mass cores 
are already 
sufficiently massive at birth to supply the entire mass of the subsequent high-mass star via local
accretion.
Thus, in ``turbulent core accretion,'' a high mass star begins its life as a cold,
virialized, high-mass core, and its subsequent evolution proceeds locally, essentially in isolation from
the surrounding clump.

The ``competitive accretion'' model explicitly relies on the funneling of gas into the center of the  clump in order to provide fresh material for the central cores to accrete enough mass to build up a high-mass star.  In other words, this model predicts that the clump must be gravitationally collapsing.  Although the ``turbulent core accretion'' model has no such explicit requirement, it stands to reason that the formation of the parsec scale dense clump itself almost certainly results from the gravitational collapse of a larger giant molecular cloud or a portion thereof.  Thus, in either scenario, infall motions due to gravitational collapse are expected.  The observational challenge remains to detect and quantify these collapse motions reliably.
We present here statistically-based evidence that infall motions, which we interpret as gravitational collapse, can be detected toward a large sample of dense molecular clumps.

\subsection{Detecting Collapse Motion via the Blue Asymmetry}

One of the main methods of detecting gravitational collapse in dense molecular clouds is to search for ``blue  asymmetries'' in the line profiles of optically thick spectral lines (see review by \citealt{Evans2003}).  The technique was first suggested by \cite{SnellLoren1977}, who noticed that the CO $1-0$ line profiles toward four dense molecular clouds (Mon R2, W3, NGC1333, and $\rho$ Ophiucus) were asymmetric.  The CO lines had two velocity peaks, and the peak with the more negative (``bluer'') radial velocity was brighter than that of the more positive (``redder'') radial velocity. \cite{SnellLoren1977} interpreted this profile as indicating the collapse of a molecular core with a radial temperature gradient, with larger temperatures toward the center.  In a collapsing cloud, an optically thick line at redshifted velocities will probe different regions than at blueshifted velocities due to the fact that the line of sight extending from the observer to the cloud first encounters redshifted velocities in the outer part of the cloud but blueshifted velocities in the inner part of the cloud.  If the excitation temperature is larger in the inner regions, an optically thick line will therefore tend to be brighter at blue velocities and fainter at red velocities.  The exact shape of the line profile depends on the density, temperature, and velocity profiles of the clump.  (See \citealt{Evans2003} for details.)

Since this pioneering work, such  blue asymmetric line profiles have been detected and interpreted as indicating collapse toward a large number of star forming regions.  For example, blue asymmetric profiles were detected toward many cores and clouds associated with low-mass star formation, such as the Bok globule B335 \citep{Zhou1993}, several ``Class 0" and ``Class 1" low-mass young stellar objects \citep{Gregersen1997, Mardones1997}, ``starless'' cores \citep{Lee1999},  and several cores in Perseus \citep{Campbell2016}.  The blue asymmetry has also been detected toward  the intermediate mass source IRAS 16293$-$2422 \citep{Walker1986}.   More recently, attempts have been made, with moderate success, to detect blue asymmetries toward high-mass  star forming regions, such as High Mass Protostellar Objects \citep{Fuller2005},
high-mass cluster-forming clumps \citep{Lopez2010},  Extended Green Objects \citep{Chen2010}, 70 \um\, dark clumps \citep{Traficante2018}, and a 17.6 \Msun\, core in the 70 \um\, dark Infrared Dark Cloud G331.372-00.116  \citep{Contreras2018}.  In these studies, the optically thick probe is typically a spectral line of a relatively abundant molecule with a high critical density $ (n_{crit} > 10^4$ \cmc), such as HCO$^+$ $1-0$, $3-2$, and $4-3$, CS $2-1$, or H$_2$CO $2_{12} - 1_{11}$ and  $3_{12} - 2_{11}$.  

While the blue asymmetry is often detected toward star-forming regions, it is by no means always detected.  Indeed, most surveys find only slightly more blue asymmetric profiles than red asymmetric profiles.  For example, in the \cite{Fuller2005} study of High Mass Protostellar Objects, only 15\% more sources show a significant blue asymmetry rather than a red asymmetry in the \hcop\, line. Moreover, \cite{Velusamy2008} suggest a bimodal distribution of asymmetries for starless cores in Orion, with roughly equal numbers of red and blue asymmetric profiles of \hcopnt\, $3-2$  in their sample.  Furthermore, in some cases the same source shows a blue asymmetry in one line and no asymmetry or a red asymmetry in another (e.g., \citealt{Walker1986}).  Thus, the observational evidence for collapse based on the blue asymmetry is often weak or ambiguous.  It is not yet established by previous work  that the ensemble of high-mass star forming clumps demonstrates definitive evidence for collapse.  

Even when the blue asymmetry is present, quantifying the degree of asymmetry is non-standard and varies from study to study.  The prototypical blue asymmetry profiles often resemble two superposed Gaussian profiles, with the blue-shifted Gaussian having the larger amplitude.  Consequently, profiles are often fitted with two Gaussians, and the parameters arising from this two-Gaussian decomposition are used to quantify the asymmetry, such as the ratio of the amplitudes of the ``blue'' and ``red'' fit Gaussians \citep{Gregersen1997}, or the velocity difference of the two Gaussians normalized by the FWHM  linewidth of a Gaussian fit to an optically thin molecular line \citep{Fuller2005}.  Another approach is to treat both the optically thick and optically thin lines as single Gaussian profiles, and to quantify the asymmetry as the difference in velocities between the Gaussian fits to the optically thick and thin lines (e.g., \citealt{Rathborne2016}), sometimes normalized by the linewidth of the optically thin line (the so-called ``$\delta v$'' parameter of  \citealt{Mardones1997}).

\subsection{Other Techniques to Infer Gravitational Collapse}

In addition to the blue asymmetry technique to infer gravitational collapse, other techniques are often employed.  In cases where the clump or core contains a central continuum source, such as radio free-free emission from an embedded H II region or thermal dust emission from a central Young Stellar Object, blueshifted emission coupled with redshifted absorption against the continuum source, also known as an ``inverse P Cygni profile,'' indicates collapse.  The presence of inverse P Cygni profiles, often inferred with interferometers, has been suggested to indicate collapse toward several high-mass star-forming regions, such as G10.6-0.4 \citep{HoHaschick1986}, G34.3+0.2, W3(OH) \citep{Keto1987}, W51 \citep{Rudolph1990}, and W49 \citep{Welch1987}. \cite{Wyrowski2016} have also detected red-shifted absorption against the continuum in a far-infrared line of NH$_3$ toward several high-mass star-forming regions using the SOFIA airborne telescope.

In addition to the search for inverse P Cygni profiles,  an alternate approach called the ``blue bulge'' technique uses spatially resolved images of Gaussian fits to spectral lines to find connected central regions of predominantly ``blue'' profiles toward the center and ``red'' profiles in the outskirts.  This technique has the advantage of finding collapse motions in the presence of rotational motions, and has been used successfully toward IRAS 16293$-$2422 \citep{Narayanan1998,NarayananWalker1998}.

As with the blue asymmetry technique, neither of these techniques is a definitive indicator of collapse.  For example, in the inverse P Cygni technique, it is sometimes unclear whether the absorbing gas is in fact associated with the clump rather than with some unrelated foreground source or sources.  Toward W49, for example, there are three distinct absorption features in the \hcop\, line, only one of which was interpreted by \cite{Welch1987} to be associated with the collapse of the clump.  The ``blue bulge'' method faces similar issues in the presence of superposed clumps or other bulk motions, such as shear.

\subsection{Theoretical Expectations}

The theory of gravitational collapse and its resulting spectral signatures in the study of star formation have a rich history.   \cite{Shu1977} proposed a self-similar ``inside-out'' collapse theory in which the inner portions of a star-forming core, initially modeled as an isothermal sphere, collapse first, resulting in a rarefaction wave propagating outward at the sound speed.  Since then, several studies have attempted to model the appearance of the spectral line emission emerging from collapsing cores or clumps, especially to try to match the observed self-reversed, blue asymmetric line profiles (e.g., \citealt{LeungBrown1977, Walker1986, Zhou1992, Myers1996, NarayananWalker1998}).  \citet{Myers2005} also explored the effects of various initial density profiles and both spherical and cylindrical geometries.
In these studies, blue asymmetric profiles are commonly produced in idealized spherical collapse models of low-mass cores where turbulent motions are small.

Specific theoretical predictions for asymmetric line shapes in collapsing high-mass star-forming regions have only recently been investigated.  Since high-mass star-forming regions have larger turbulent linewidths than low-mass star-forming regions (e.g., \citealt{Rathborne2016}), collapse motions might be expected to be more difficult to detect. The recent realization of the importance of filamentary structures and turbulent motions in the formation of high-mass stars and star clusters requires more sophisticated modeling that includes these more complex geometries and more realistic motions coupled with detailed radiative transfer calculations \citep{Smith2012, Smith2013, Chira2014}.   These models show that the blue asymmetry is indeed often difficult to detect from collapsing clumps in a turbulent, filamentary environment.  Because the motions are complex and multiple regions can overlap along the line of sight, the observed line profiles from optically thick molecular tracers vary significantly when viewed from different angles \citep{Smith2012}.  Moreover, the degree of asymmetry for \hcop\, lines from collapsing high mass clumps is predicted to be small compared to typical linewidths.  Indeed, in the \cite{Smith2013} simulations the \hcop\, profile is shifted only slightly toward the blue;  in only half the cases is the simulated velocity shift toward collapsing regions $>0.5$ \kms.  
 
To summarize, theory suggests that collapse motions in high-mass star-forming regions are expected to result in blue asymmetries, but the asymmetry is expected to be small and often diminished by the effects of viewing angle and complex motions and structure.  For this reason, the observational detection of this signal toward dense clumps would benefit from averaging over a large sample in order to obviate the effects of random viewing angles and turbulent fluctuations.  To that end, we have analyzed data from the MALT90 survey \citep{Foster2011, Foster2013, Jackson2013}, which has obtained  high critical density molecular line data for the largest number of clumps to date.  We employ the \hcop\, line as our optically thick probe with which to search for blue asymmetries, and the \nthp\, line as our optically thin probe to establish the clumps' systemic velocities. 

\subsection{A New Asymmetry Parameter $A$}

The MALT90 spectra for the most optically thick species exhibit a wide variety of line shapes, with some showing evidence of outflows indicated by red or blue line wings or both, as well as absorption dips of various depths. Specifically, the optically thick \hcop\, lines used for our analysis often have complex lineshapes for which neither single- nor double-Gaussians provide good fits to the data  (see Figure \ref{fig:fig1}). [The hyperfine structure of  the optically thick \hcn\, line produces very complex line shapes that preclude its use in this type of analysis.]  Thus, the choice of a more robust quantity to indicate a line's asymmetry might benefit from abandoning specific assumptions about the intrinsic shape of the line profiles.  

 In order to quantify the degree of line asymmetry, this paper employs a new, simple measure of the asymmetry that is independent of any assumption about line shape.  This approach is similar to the use of the line ``skewness'' from \cite{Gregersen1997}, but is simpler.  Our technique is to establish the systemic velocity with an optically thin line, and then to measure the degree of asymmetry of the optically thick line by the quanitity  $A = {{I_{blue}-I_{red}}\over{I_{blue}+I_{red}}}$, where $I_{blue}$ represents the integral of the line profile blueward of the systemic velocity and $I_{red}$ the integral redward of the systemic velocity. $A$ represents the fraction of the total line flux that lies to the blueshifted side of the systemic velocity.   In this scheme, line profiles with a blue asymmetry will have values of $A >0$, symmetric line profiles will have $A=0$, and line profiles with a red asymmetry will have $A <0$.  Larger absolute values of $A$ are more asymmetric.  In the absence of absorption against a continuum source, $A$ is bounded between $-1$ and $+1$.

\subsection{Specific Predictions}
   While the great variety of line shapes combined with the statistical uncertainties in the measurements will lead to broad distributions in the asymmetry parameter $A$, if on average the ensemble of clumps is in collapse, theory predicts that the distribution in $A$\, should be shifted toward positive values:  both the mean and the median of the distribution of $A$ should be greater than zero, and the fraction of sources with a positive asymmetry should be greater than 50\%.  

The simulations of \cite{Smith2013} suggest a small predicted value for $A$.  For the optically thick \hcop\, line, these simulations obtain a median velocity shift $V_{shift}$ of $\sim$ 0.5 \kms.  Thus, as a rough approximation we can estimate $A \sim V_{shift}/\Delta V$, which for the sample's mean \hcop\, linewidth $\Delta V \sim 5$ \kms\, (see  Figure \ref{fig:fig14} in the Appendix) corresponds to a predicted value of $A \sim 0.1$.  To detect such a small shift is observationally challenging and requires both a large sample and a careful analysis of the observational and systematic errors.

Another rudimentary implication arises if we imagine at the outset a simple Gaussian lineshape that is modified by enhanced emission on the blue side and/or absorption on the red side.    We would expect that the peak moves to the blue side, so that the local slope of the primary line spectrum at the systemic velocity becomes negative.  We tested this idea by fitting lines to the \hcopnt\, spectrum through the region $\pm$5 spectral channels ($\sim \pm$0.6 \kms ) about the \nthpnt\, reference velocity.  As with the asymmetries, we expect wide distributions in the slopes of these line segments, but the mean and median observed for the ensemble should be negative, and the fraction of sources with negative slopes should be greater than 50\%.

These, then, are three key predictions for the analysis of asymmetric profiles for an optically thick tracer of an ensemble of clumps that are undergoing gravitational collapse: (1) a positive value of the mean and median of the enesemble distribution of the asymmetry parameter $A$, (2)  a negative value of the mean and median of the ensemble distribution of the local slope of the line profile at the systemic velocity, and (3) a larger fraction of sources showing parameter values consistent with collapse than those inconsistent with collapse, namely positive values of $A$ and negative values of the local slope of the line profile at the systemic velocity.

\section{Observations}
\label{sec:observations}
This study employs the molecular line spectra observed in the MALT90 survey \citep{Foster2011, Foster2013, Jackson2013} and described by \citet{Rathborne2016}.  MALT90 used the ATLASGAL 870 \um\, survey \citep{Schuller2009} to select targets identified in a compact source catalog \citep{Contreras2013} likely to be dense clumps, and then used the ATNF Mopra 22-m telescope to map a $4' \times 4'$ region around these targets; the central  $3' \times 3'$ portion of each map has superior noise characteristics due to the on-the-fly mapping process employed by MALT90.  The pixel size is $9''$, the angular resolution is $38''$, and the spectral resolution is 0.11 \kms.  Additional ATLASGAL targets falling within the mapped regions were added to our source list:  in a total of 2014 MALT90 maps we identified a total of 3246 ATLASGAL targets.

 For each target, we calculated for each of the sixteen observed spectral lines a spectrum  averaged over a $3 \times 3$ pixel block centered on the peak of the 870 $\mu$m dust continuum emission observed by ATLASGAL.  The median 1$\sigma$ brightness sensitivity of the averaged spectrum is 0.18 K on the $T_A^*$ scale.  Approximately 5 percent of the targets (175 out of 3246) had no line signal passing our detection threshold.   In approximately  9 percent of the targets we found two distinct velocity components separated by more than 15 \kms\, resulting in two distinct MALT90 source entries in the MALT90 catalog.  The final catalog contains 3556 source entries.  

For this current study, in order to exclude the extreme clouds in the Galactic Center with very broad linewidths, we have limited the Galactic longitude of the sources to $295^{\circ} < l < 350^{\circ}$, which includes 2029 sources.  Further, in order to avoid confusion arising from two or more clouds along the same line of sight, we have retained  sources  with only a single velocity component, i.e., those sources with a catalog suffix ``$\_ S$" as described in  \citet{Rathborne2016}; this reduces the number of sources to 1807.  Finally, for this same reason, we have also rejected sources  for which our line-fitting analysis detected a second velocity component or non-Gaussian wings for \nthpnt\, within 15 \kms , i.e., those with the NG flag set to 2 as discussed in  \citet{Rathborne2016}; this resulted in a final selection of 1794 sources  to be included in the asymmetry analysis.  In this analysis we require the integrated intensity signal-to-noise ratio for all lines to exceed a set minimum, typically 4.0, leading to the reported number of 1,093 sources in the main analysis described below.

\section{Analysis}

We calculate the blue/red asymmetry parameter $A$ of a ``primary'' spectral line with respect to a reference velocity derived from a ``reference'' line: $$A = {{I_{blue}-I_{red}}\over{I_{blue}+I_{red}}}~~,$$ where $I_{blue}$ is the integrated intensity at velocities less than the reference velocity and $I_{red}$ is the integrated intensity at more positive velocities.   The integration range for $I_{blue}$ extended from  $V_{ref}-2\Delta{V}$ to $V_{ref}$, where $V_{ref}$ is the reference line velocity and $\Delta{V}$ is the observed velocity FWHM dervived from a single Gaussian fit for the primary line.  (See  \cite{Rathborne2016} for the fitting procedure and the Appendix for slight  modifications).  The integration range for $I_{red}$ mirrored this to the higher velocity side.  The intensity from the single spectral channel of the primary line within which the reference velocity fell was excluded from both the ``blue'' and ``red'' integrations.  The integration range of $2\Delta{V}$ is chosen to be large enough to include all significant emission yet small enough to exclude too many signal-free channels. The exact choice of the integration range does not greatly affect our results; our findings are robust against reasonable variations in the width of the integration range.

\subsection{Uncertainty of a measured asymmetry}
The statistical uncertainty in a measured asymmetry has two leading contributions:  the random noise in each spectral channel, and the uncertainty in the fitted reference velocity.  The first contribution to the asymmetry uncertainty was calculated by propagating the uncertainties in the Blue and Red integrated intensities, which were calculated as the product of  the measured rms noise in the spectrum, the square root of the number of channels in each integration range, and the velocity width of a spectral channel 
($\sigma_I = \sigma_{T_A} \sqrt{N_{channels}} \delta V$).  We estimate  the second contribution by calculating the rms deviation of an ensemble of asymmetry values computed while dithering the reference velocity from its nominal value by an offset that was normally distributed with zero mean and a dispersion equal to the reference velocity uncertainty as returned by the \cite{Rathborne2016} fitting procedure.  Our  fit to the \nthpnt\, spectrum is discussed in detail in the Appendix.  We combined the two uncertainty contributions in quadrature to estimate the total statistical uncertainty in each asymmetry measurement.  Our estimates of the systematic uncertainties in the asymmetry measurements are discussed in the Appendix.

\section{Results}
\subsection{Asymmetry Parameter $A$}
\begin{figure}
\begin{center}
\includegraphics[scale=0.751, angle=0]{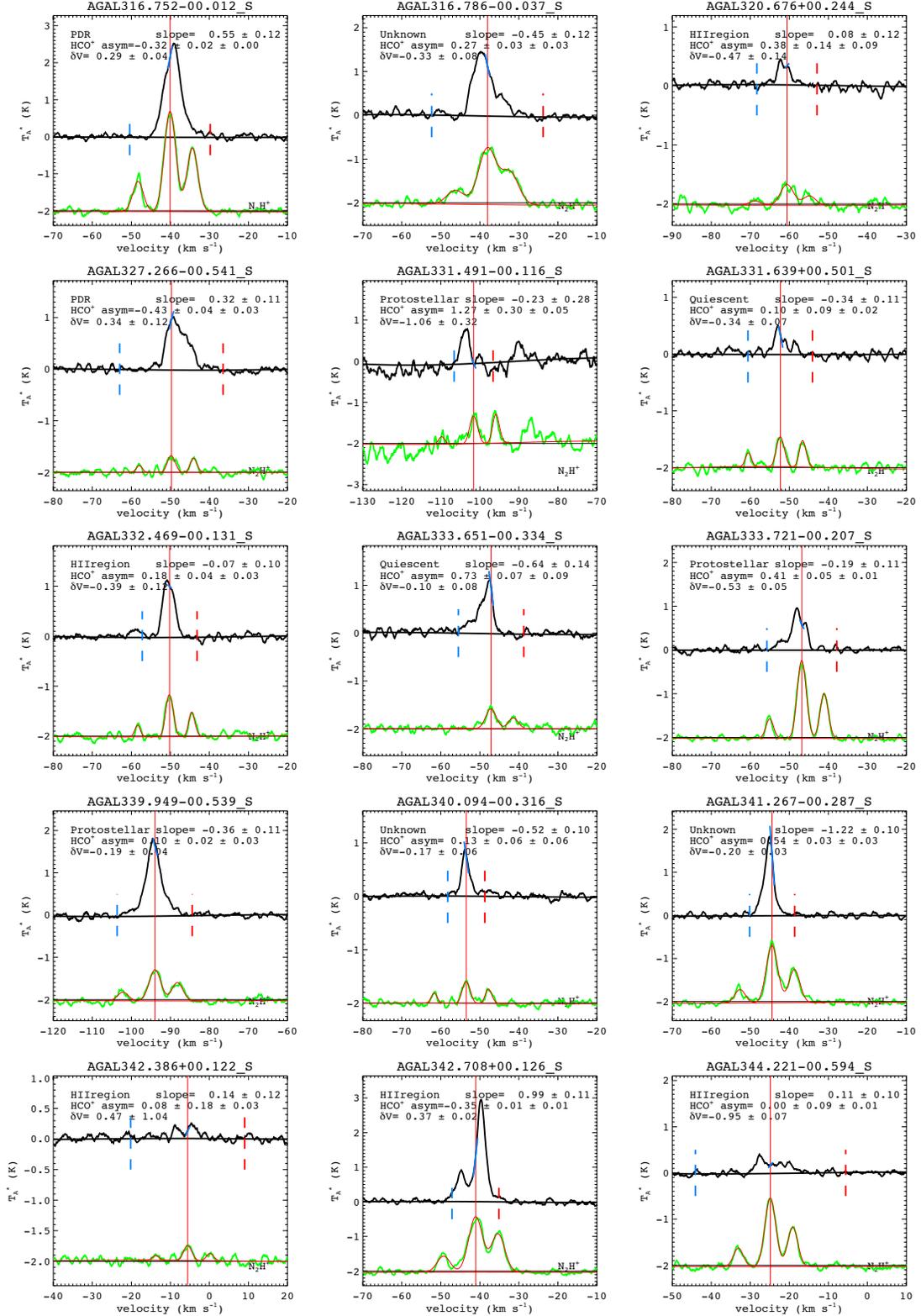} 
\caption{ Spectra of a sample of MALT90 sources.  The lower trace shows the \nthpnt\, spectrum in green, and the fit to that spectrum in red.   The vertical red line marks the velocity fitted to the \nthpnt\, spectrum.  The upper trace shows the \hcopnt\, spectrum; the vertical blue and red dashed lines show the left and right boundaries of the Blue and Red integration regions respectively.  The short blue line segments show the local fit to the \hcopnt\, spectrum at the \nthpnt\, velocity. }
\label{fig:fig1}
\end{center}
\end{figure}

Typically,  \hcop\, (hereafter ``\hcopnt'') will have the greatest optical depth of the lines observed in MALT90 (e.g., \citealt{Sanhueza2012, Hoq2013}), and consequently will provide the clearest infall signatures.   We use \nthp\, (hereafter ``\nthpnt'') as the ``reference line'' to determine the systemic velocity, as this species is expected to be optically thin (an assumption we will address and verify below; see also \citealt {Sanhueza2012, Hoq2013}), and its high detection rate \citep{Rathborne2016} will provide the largest number of sources for the asymmetry analysis.  Figure \ref{fig:fig1} shows the \hcopnt\, and \nthpnt\, spectra for a random selection of MALT90 sources.   For visual clarity these spectra have been smoothed by a Savitzky-Golay filter of second order with a width of 21 channels, which reduces the noise by a factor of 0.34; all the fitting and analysis were done at the original, unsmoothed spectral resolution of 0.11 \kms.  Figure \ref{fig:fig1} also shows (in blue on the \hcopnt\, spectrum) the fitted line segments discussed above.  The wide variety of \hcopnt\, line shapes illustrates the diversity of physical properties and internal strcutures of the MALT90 sources and the challenge of extracting a general infall signature on a source by source basis.  Nevertheless, an initial analysis of MALT90 data hinted at the presence of a {\it statistical} collapse signal in the ensemble of  MALT90 clumps; \cite{Rathborne2016} find that Gaussian fits to the line profiles on average show slightly smaller (bluer) velocities in \hcopnt\, than in \nthpnt.

\begin{figure}
\begin{center}
\includegraphics[scale=1.0, angle=0]{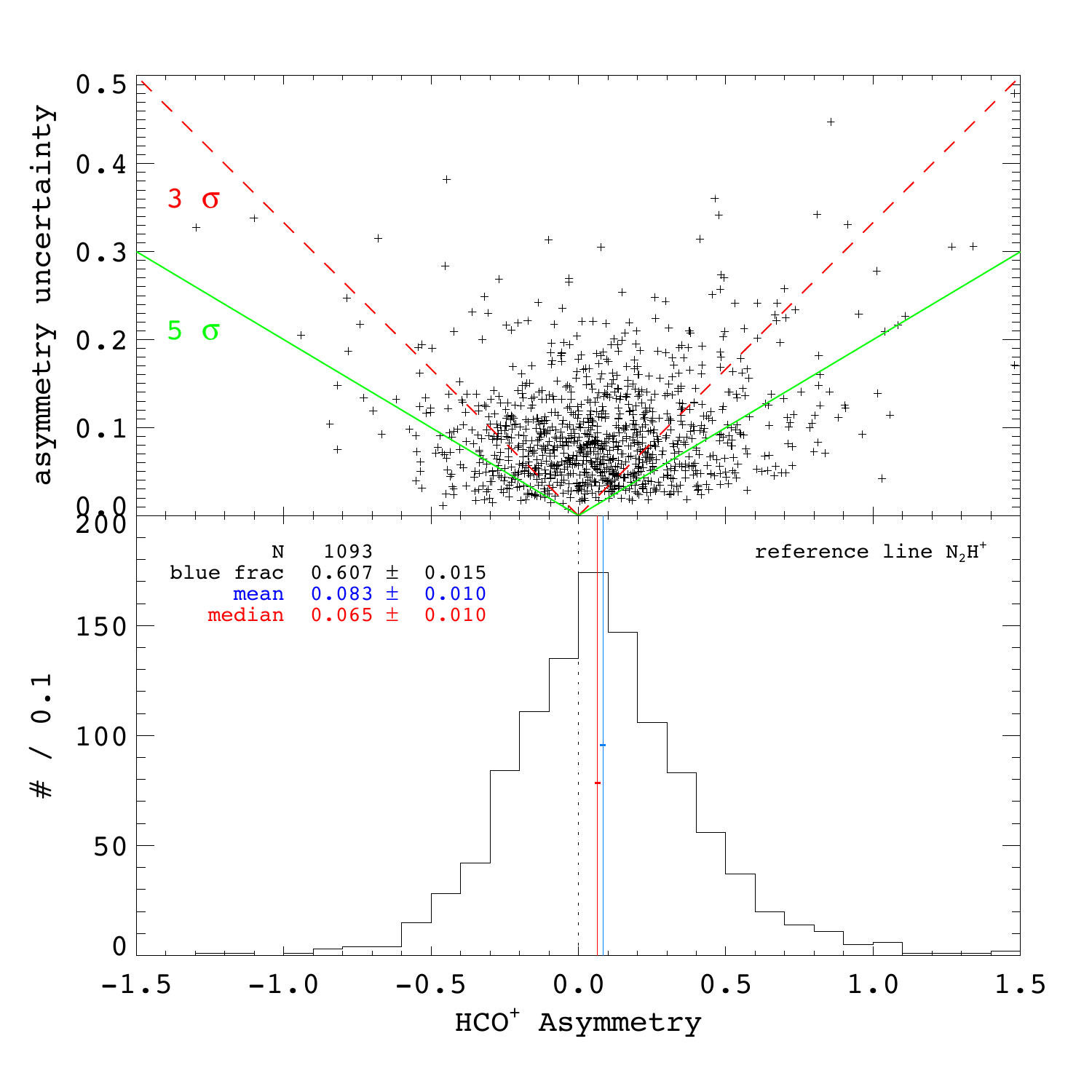} 
\caption{The asymmetry parameter $A$ for \hcopnt , with \nthpnt\, as the reference line.   The upper panel shows the asymmetries on the x-axis and the 1$\sigma$ uncertainty in the asymmetry on the y-axis.  Points lying below the dashed red and solid green lines have an asymmetry value that is at least three or five times its uncertainty respectively.  The lower panel shows a histogram of the asymmetry. The vertical red and blue lines and text denote the median and mean of the distribution; the horizontal red and blue lines indicate the $\pm 1 \sigma$ uncertainties in the median and mean. }
\label{fig:fig2}
\end{center}
\end{figure}

In Figure \ref{fig:fig2} we present the asymmetry measurements for the 1093 MALT90 sources passing our selection criteria and having an integrated intensity signal-to-noise ratio of at least 4.0 in both the \hcopnt\, and \nthpnt\, lines.  The lower panel shows a histogram of the asymmetry parameter $A$, which exhibits an excess of positive asymmetries.  The fraction of sources with positive asymmetries is $0.607 \pm 0.015$. The blue and red vertical lines and text indicate the mean and median of the asymmetry distribution.  The uncertainties in the mean and median have been calculated as the rms deviations of values extracted from an ensemble of asymmetry distributions generated from the observed distribution by bootstrap resampling with replacement \citep[see e.g.][]{Simpson1986}.  

For our sample of 1093 clumps, we find a broad distribution in $A$ with a mean value of $ 0.083 \pm 0.010$ and a median value of $0.065 \pm 0.009$.  This is the central result of this study.  The positive values of the mean and median are highly statistically significant and provide statistical evidence for collapse in this ensemble of sources.  We note that there are both positive and negative asymmetries with high statistical significance, consistent with theoretical expectations and simulations of cores evolving in turbulent, filamentary environments \citep{Smith2012, Smith2013, Chira2014}.

To test the sensitivity of our measurement of $A$ to a possible systematic uncertainty in the \nthpnt\, rest frequency, we repeated the analysis while offsetting the \nthpnt\, velocity by $\pm$ 0.03 \kms, a reasonable estimate of the maximum experimental and systematic uncertainty  (see the Appendix for discussion).  The results are shown in Table \ref{table:syserrs}.  Our finding of an overall positive ensemble value for $A$ for this sample of clumps is robust against the estimated systematic errors.

\begin{figure}
\begin{center}
\includegraphics[scale=1.0, angle=0]{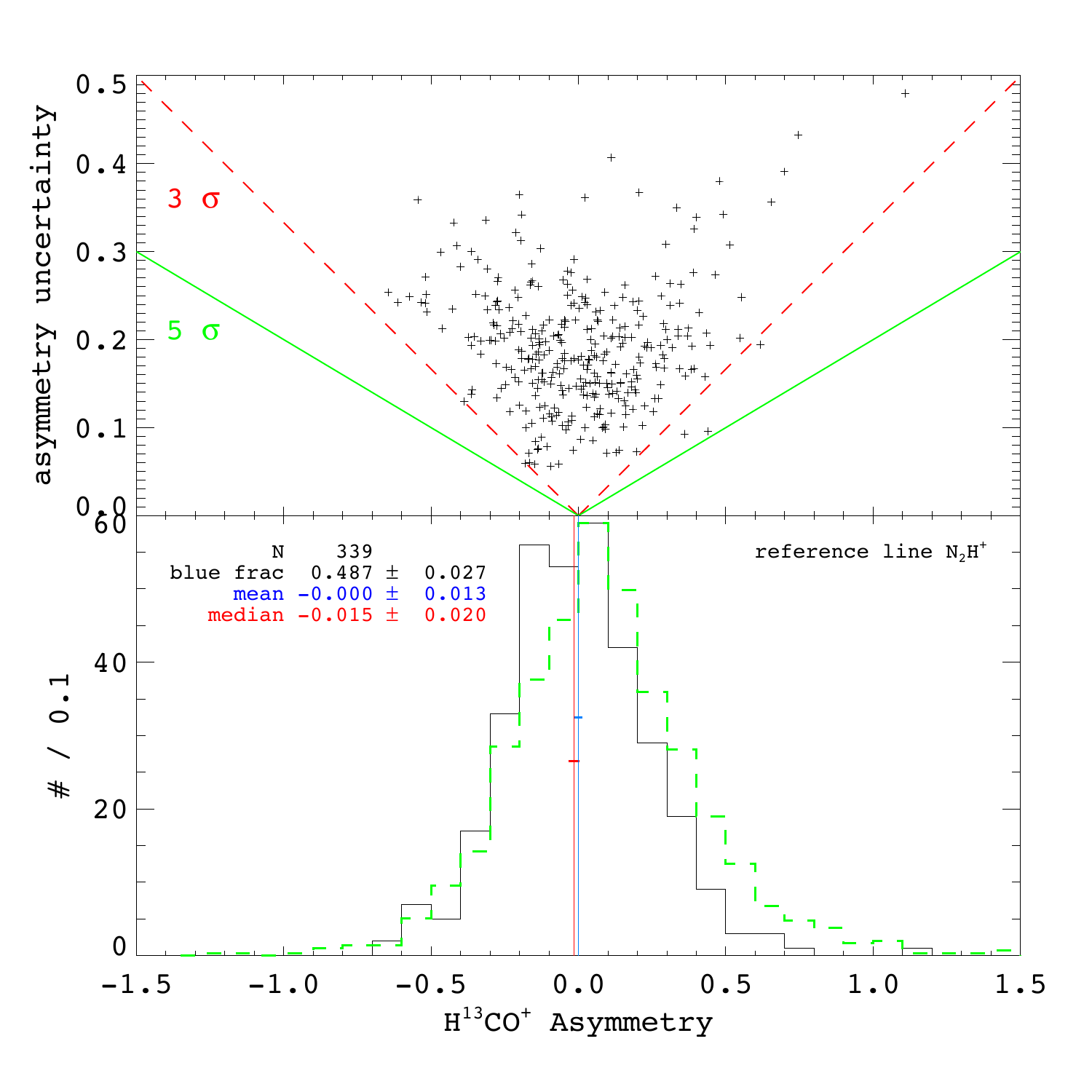} 
\caption{The asymmetry parameter $A$ for \htcopnt , with \nthpnt\, as the reference line.The upper panel shows the asymmetries on the x-axis and the 1$\sigma$ uncertainty in the asymmetry on the y-axis.  Points lying below the dashed red and solid green lines have an asymmetry value that is at least three or five times its uncertainty respectively.  The lower panel shows a histogram of the asymmetry. The vertical red and blue lines and text denote the median and mean of the distribution; the horizontal red and blue lines indicate the $\pm 1 \sigma$ uncertainties in the median and mean.  The distribution of $A$ for \hcopnt, normalized to have the same peak, is superposed in green.  Note the clear shift to more positive values for \hcopnt.  A K-S test gives a probability of $9\times10^{-5}$ that these two distributions were drawn from the same parent population. }
\label{fig:fig3}
\end{center}
\end{figure}

Since \htcopnt\, is less abundant than \hcopnt\, by a factor of $\sim$20 to 60 \citep{Stark1981},  \htcop\, (hereafter ``\htcopnt'') should almost always be optically thin.  Consequently we expect no indication of collapse if we examine the asymmetry of \htcopnt, again using \nthpnt\, as a reference.  Our analysis of the \htcopnt\, asymmetry parameter $A$ is illustrated in Figure \ref{fig:fig3}. Of the 390 sources with the \htcop\, signal-to-noise ratio greater than 4.0, both the mean  ($0.000 \pm 0.014$) and the median ($-0.015 \pm 0.020$) of $A$ are consistent within the errors with a value $A=0$, and the positive fraction is 0.487 $\pm$ 0.027, consistent with 50\%,  as expected. 

\begin{figure}
\begin{center}
\includegraphics[scale=1.0, angle=0]{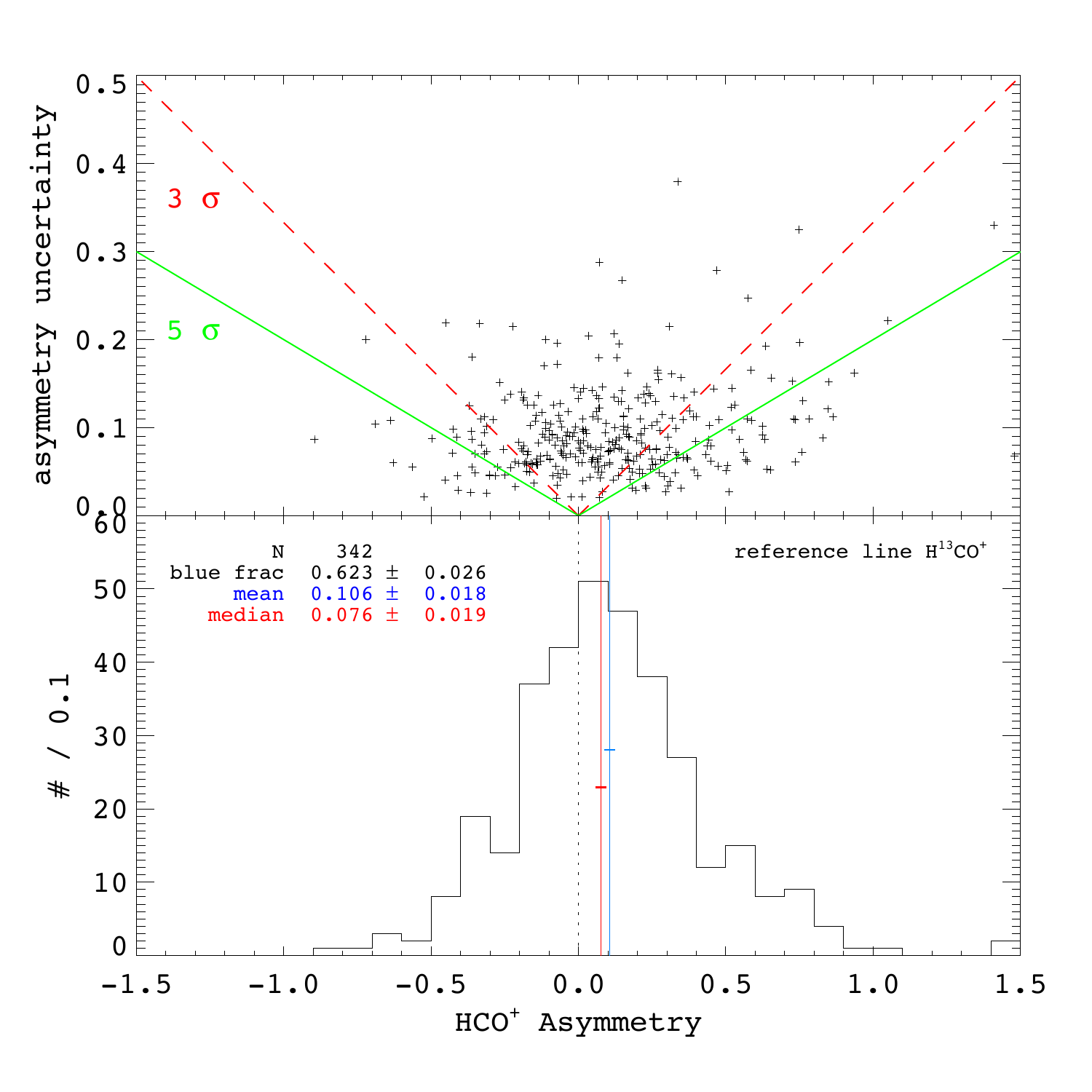} 
\caption{The asymmetry parameter $A$ for \hcopnt , with \htcopnt\, as the reference line. The upper panel shows the asymmetries on the x-axis and the 1$\sigma$ uncertainty in the asymmetry on the y-axis.  Points lying below the dashed red and solid green lines have an asymmetry value that is at least three or five times its uncertainty respectively.  The lower panel shows a histogram of the asymmetry. The vertical red and blue lines and text denote the median and mean of the distribution; the horizontal red and blue lines indicate the $\pm 1 \sigma$ uncertainties in the median and mean. }
\label{fig:fig4}
\end{center}
\end{figure}

We expect that when  \htcopnt\, is detected with $4 \sigma$ significance,  \hcopnt\, will be very optically thick; this should result in increased asymmetry signatures.  This expectation is borne out in Figure \ref{fig:fig4} where we exhibit the asymmetry measurements for \hcopnt\, with \htcopnt\, as the reference line.  Indeed, when \htcopnt\, is used as the reference line, the mean of $A$ ($0.106 \pm 0.012$), the median of $A$ ($0.076 \pm 0.010$), and the fraction of blue sources ($0.623 \pm 0.026$) are all larger than when \nthpnt\, is used as the reference line.

\subsection{Slope at Reference Velocity} 

\begin{figure}
\begin{center}
\includegraphics[scale=1.0, angle=0]{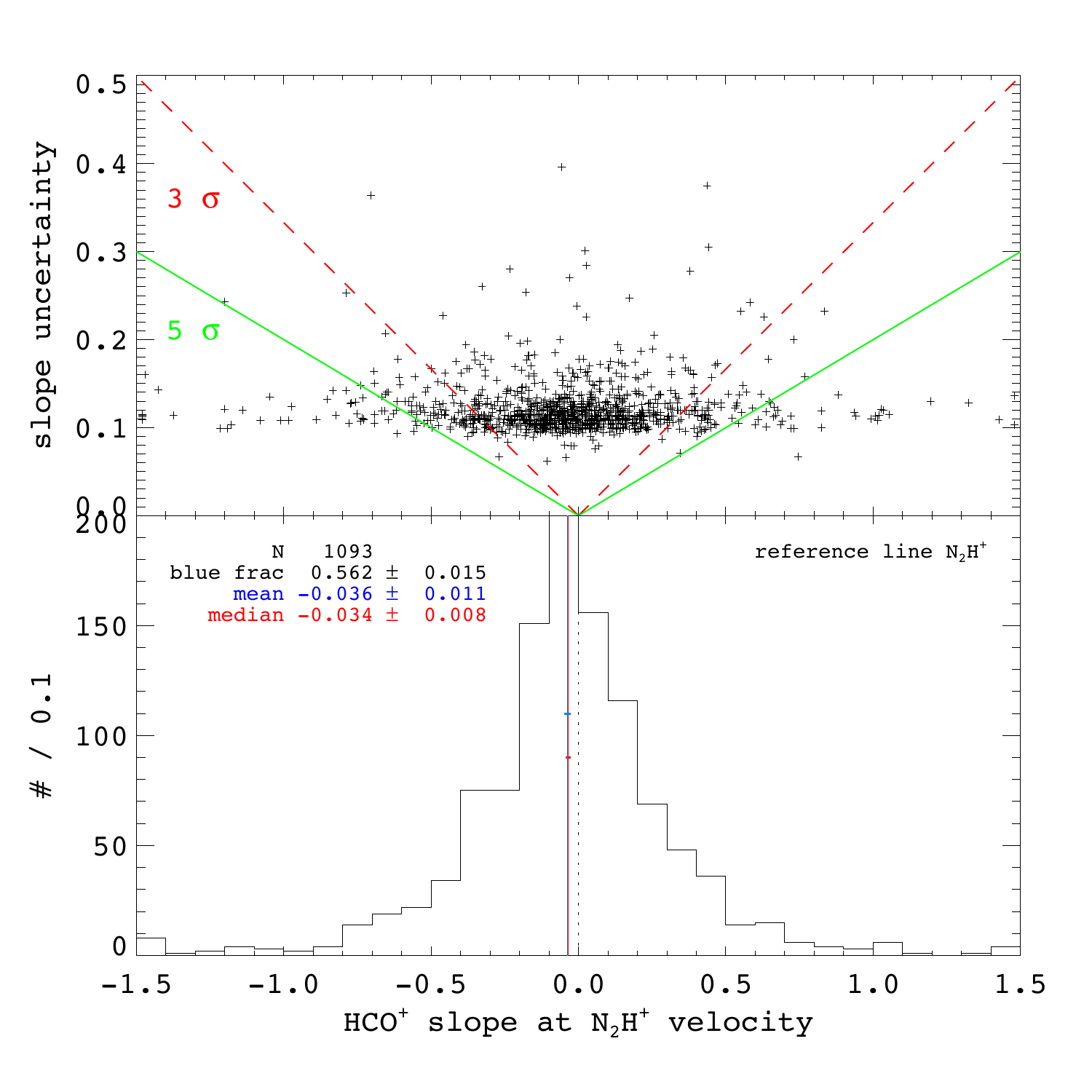} 
\caption{The local slope of the \hcopnt\, line at the systemic velocity determined by \nthpnt.  The slope is $d(T_A^*)/d(V_{LSR})$
 in units of K km$^{-1}$ s.  The upper panel shows the slopes on the x-axis and the 1$\sigma$ uncertainty in the asymmetry on the y-axis.  Points lying below the dashed red and solid green lines have an asymmetry value that is at least three or five times its uncertainty respectively.  The lower panel shows a histogram of the slopes. The vertical red and blue lines and text denote the median and mean of the distribution; the horizontal red and blue lines indicate the $\pm 1 \sigma$ uncertainties in the median and mean. }
\label{fig:fig5}
\end{center}
\end{figure}

\begin{figure}
\begin{center}
\includegraphics[scale=1.0, angle=0]{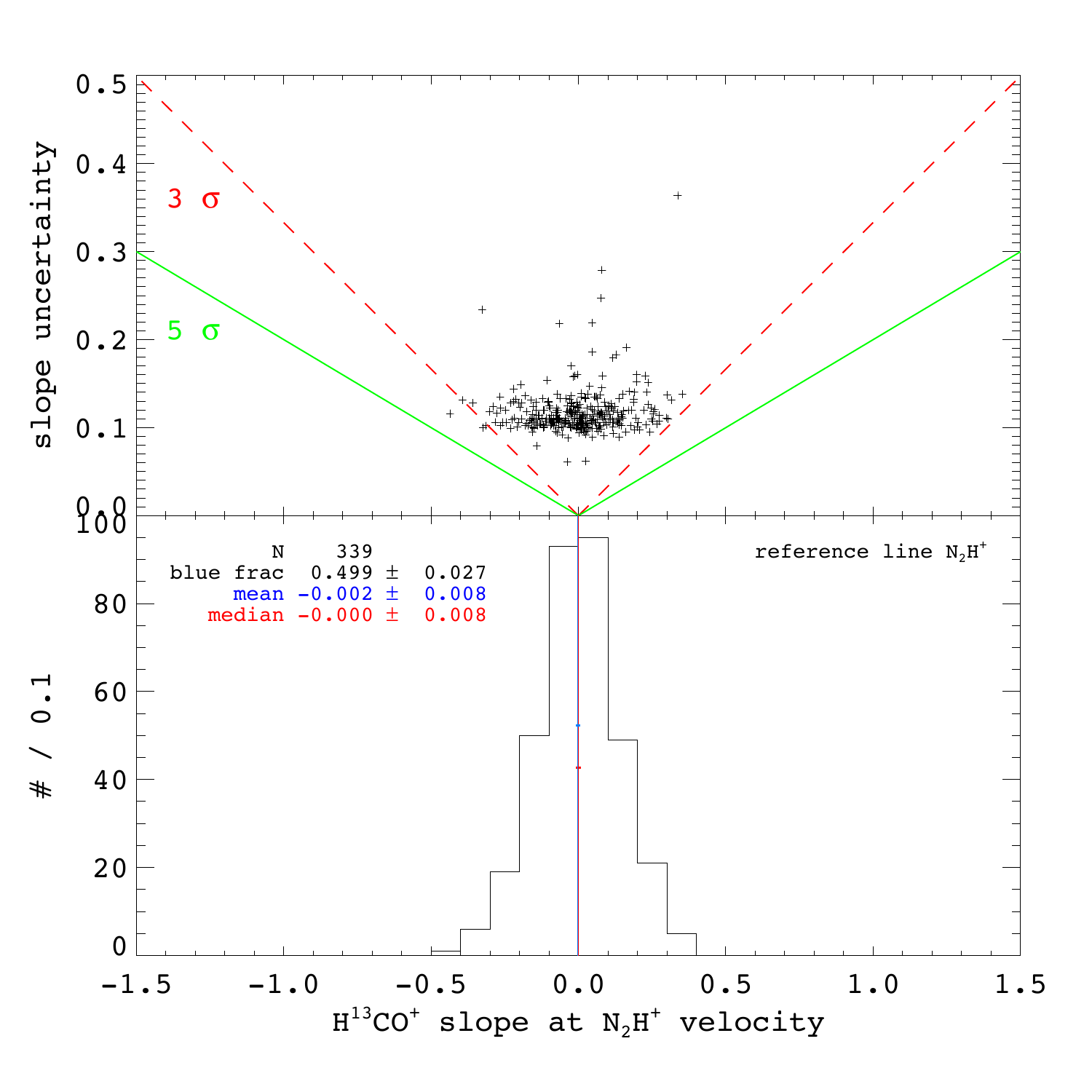} 
\caption{The local slope of the \htcopnt\, line at the systemic velocity determined by \nthpnt.  The slope is $d(T_A^*)/d(V_{LSR})$
 in units of K km$^{-1}$ s.  The upper panel shows the slopes on the x-axis and the 1$\sigma$ uncertainty in the asymmetry on the y-axis.  Points lying below the dashed red and solid green lines have an asymmetry value that is at least three or five times its uncertainty respectively.  The lower panel shows a histogram of the slopes. The vertical red and blue lines and text denote the median and mean of the distribution; the horizontal red and blue lines indicate the $\pm 1 \sigma$ uncertainties in the median and mean. }
\label{fig:fig6}
\end{center}
\end{figure}

As mentioned above, a second, independent test for the presence of a blue asymmetry is to measure the local derivative of the optically thick line profile at the systemic velocity.
Figure \ref{fig:fig5} and \ref{fig:fig6} show this analysis.  In Figure \ref{fig:fig5} the local slope $d(T_A^*)/d(V_{LSR})$ of \hcopnt\, at the systemic velocity as determined by \nthpnt\, is shown in the same format as Figures \ref{fig:fig2}, \ref{fig:fig3}, and \ref{fig:fig4}.  A small but statistically significant negative value is found for both the mean and the median of the distribution, and more than 50\% of the sample show negative values for the local slope.  In Figure \ref{fig:fig6} the slope is determined for \htcopnt\, rather than \hcopnt.  For \htcopnt, within the errors, the slope is consistent with zero, and the fraction of sources with negative slope is consistent with 50\%.

\subsection{Comparison with the \citet{Mardones1997} $\delta v$ Method}  
 
Our  asymmetry  parameter  $A$  has  the  advantages  of  being  conceptually  simple,  scale-independent,  and closely tied to the  fundamental  measured  quantity $ I(v)$.   
Like other methods, it has the potential  disadvantage in its tracing of emission from  all regions along the line of sight and the potential contamination from outflows and expansion motions of embedded \hii\, regions.  Both confusion with multiple sources and contamination of the infall signal from internal expansion motions might be expected to be more problematic in the higher mass clumps considered here.  

 As  addressed  in  the  Introduction,  other  asymmetry  measures  have  been  developed  to  infer collapse.    Our  measurement  of  the  primary  line  slope  at  the  
reference  line  velocity  is  such  a  measure.    Since other measures are in common use, it is useful to compare the results of our new $A$ parameter with another commonly used 
quantitative parameter used to infer infall, namely the $\delta v$ parameter introduced by \cite{Mardones1997} and since adopted by other studies  (e.g.,  \citealt{Velusamy2008}).

\citet{Mardones1997}  reported  a  search  for  evidence  of  infall  motions  in  a  selection  of  47  nearby    low-mass  star-forming  regions.  
 To  look  for the ``blue  asymmetry''  in  
the  spectrum  of  a  presumably  optically thick  molecular  line,  they  defined  a  parameter  $\delta  v  =  (V_{thick}-V_{thin})/\Delta  V_{thin}$,  where  $V_{thick}$
 is  the  velocity  of  the  peak  of  the  spectrum  of  the  optically  thick  line,  $V_{thin}$  is  the  velocity  of  the  optically  thin  reference  line,  
and  $\Delta  V_{thin}$  is  the  FWHM  velocity  width  for  the  thin  line.    The  optically  thick  lines  in  their  study  were  H$_2$CO $2_{12} - 1_{11}$  and  CS $(2-1)$;  
the  optically  thin  line  was  \nthp  .    
Absorption  in  the  presence  of  infall  will  shift  the  peak  of the  optically  thick  line  blueward,  resulting  in  a  negative  value  of  $\delta  v$,  while 
motion from outflows  or  rotation  would  be  expected  to  be  stochastic  in  sign  and  result  in  a  symmetric  distribution  of  $\delta  v$ with a mean or median value of zero.  

In determining $\delta v$, division  by  the  optically thin  line’s  velocity  width  is  meant  to  provide  a  
dimensionless  parameter  and  to  allow  the  comparison  of  infall  signatures  in  sources  with  different  line  widths.    The  \nthpnt\, linewidths
 in  the \citet{Mardones1997}  sample  range  from  0.3  to  1.8  \kms. In  our  sample  of  high-mass star-forming  regions,  the  \nthpnt\, linewidths  range  from  
approximately  1  to  5  \kms\,  
(as discussed  in  the  Appendix),  so  consideration  of  the $\delta v$  infall  measure  might  also  allow  comparison  of  low-mass  star  and  high-mass  star-forming  regions.  

\begin{figure}
\begin{center}
\includegraphics[scale=1.0, angle=0]{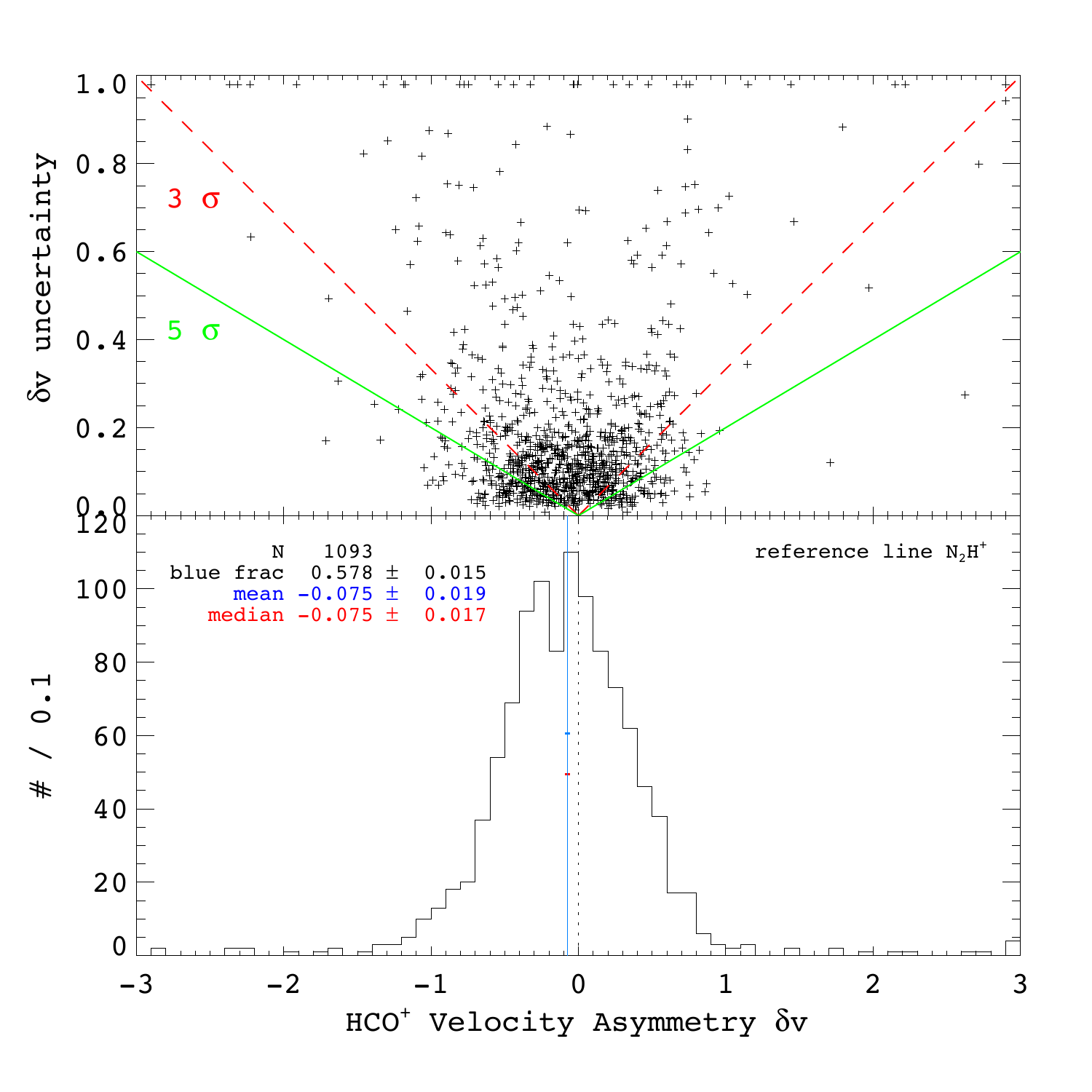} 
\caption{The $\delta v$ parameter of \citet{Mardones1997} using the peak channel of the \hcop\, line referenced to \nthp.
 The upper panel shows the values of $\delta v$ on the x-axis and the 1$\sigma$ uncertainty in the asymmetry on the y-axis.  Points lying below the dashed red and solid green lines have an asymmetry value that is at least three or five times its uncertainty respectively.  The lower panel shows a histogram of the distribution of $\delta v$. The vertical red and blue lines and text denote the median and mean of the distribution; the horizontal red and blue lines indicate the $\pm 1 \sigma$ uncertainties in the median and mean. }
\label{fig:fig7}
\end{center}
\end{figure}

\begin{figure}
\begin{center}
\includegraphics[scale=1.0, angle=0]{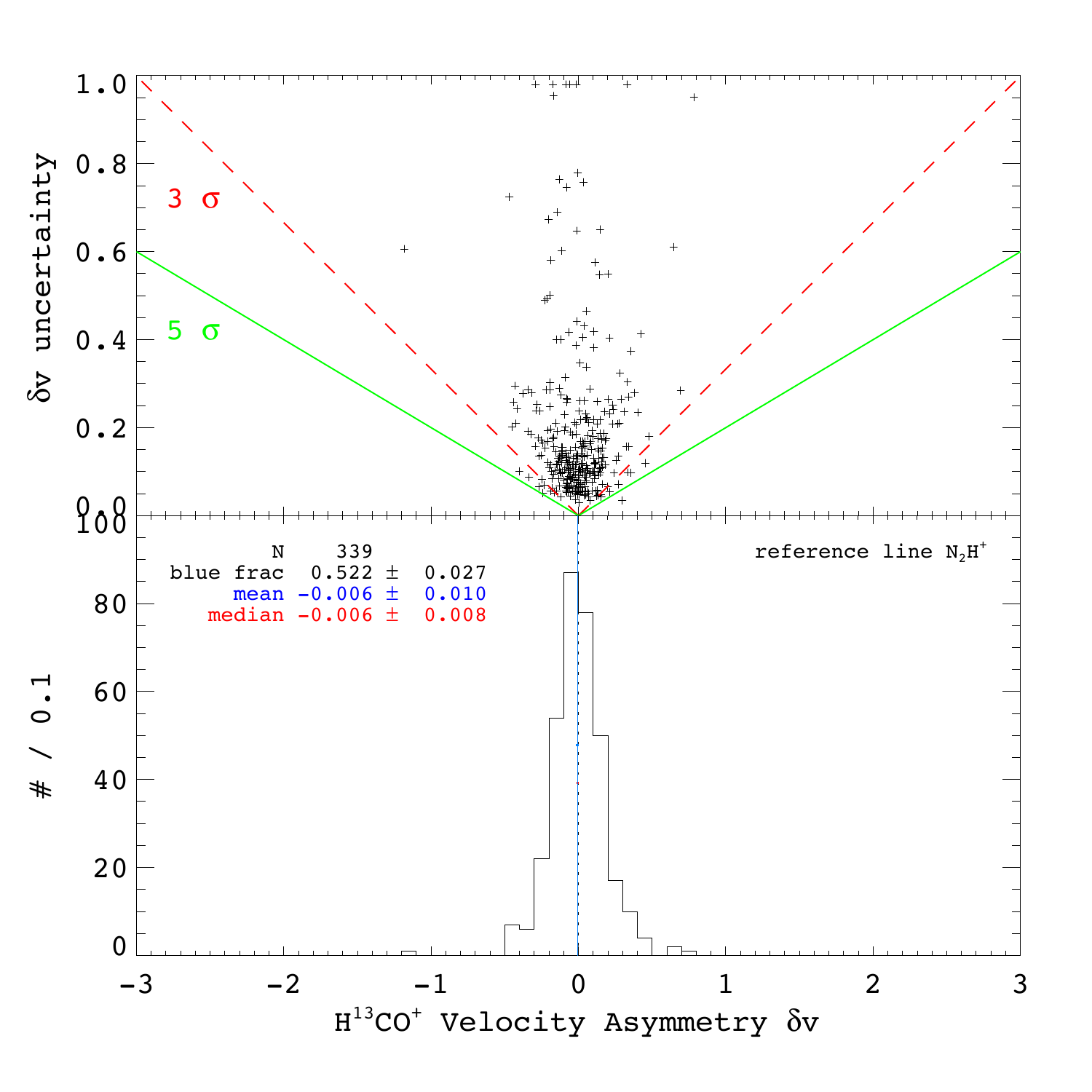} 
\caption{  The $\delta v$ parameter of \citet{Mardones1997} using the peak channel of the \htcop\, line referenced to \nthp.   
The upper panel shows the values of $\delta v$ on the x-axis and the 1$\sigma$ uncertainty in the asymmetry on the y-axis.  Points lying below the dashed red and solid green lines have an asymmetry value that is at least three or five times its uncertainty respectively.  The lower panel shows a histogram of the distribution of $\delta v$. The vertical red and blue lines and text denote the median and mean of the distribution; the horizontal red and blue lines indicate the $\pm 1 \sigma$ uncertainties in the median and mean.  }
\label{fig:fig8}
\end{center}
\end{figure}

\citet{Mardones1997}  determined  $V_{thick}$  from  a  local  Gaussian  fit  to  the  brightest  peak  of  the  optically  thick  line.    They  explored  other  
approaches,  including  using not a Gaussian fit but rather the spectral channel of the peak intensity as a measure of $V_{thick}$,  and  conclude that values of 
$V_{thick}$ using these alternate  methods  agreed  within  the rms  uncertainty  with those obtained from the Gaussian fit.
As  evident  in  the  MALT90  spectra  shown  in  Figure \ref{fig:fig1},  even  a  local  Gaussian  will  be  a  poor  approximation  to  many  of  
our  complex  line  shapes.    Further,  the  channel  width  and  noise  in  the  MALT90  data  are  larger  than  those  quantities  in  the  
spectra  of low-mass regions considered in \citet{Mardones1997},  boding  ill for  the  robustness  of  Gaussian  fits  to  just  a  few  channels.    
Instead, we determine  the  peak  velocity  $V_{thick}$ to be the velocity of the  maximum intensity  of  the  \hcopnt\,  spectrum within $\pm 2\Delta V$ of  the  reference  
velocity.    To  estimate  the  error  in  $V_{thick}$, we  created  an  ensemble of  100  test  spectra,  
each  formed  from  the  initial  \hcopnt\,  spectrum  by  adding  normally  distributed  random  noise  to  each  channel  based  on  the  rms  noise  of  the  initial  spectrum.     
 The  rms  deviation  of  the  ensemble  of  $V_{thick}$  values  derived  from  these  test  spectra  was  used  as  the  error  $\sigma(V_{thick})$.    
Using  the  errors  in  $V_{thin}$  and  $\Delta  V_{thin}$  from  our  fitting  procedure,  we  then  determined  the  error  $\sigma  (\delta  v)$.
We  performed  this  analysis  on  spectra  both  with  and  without  smoothing  as  described  for  Figure  \ref{fig:fig1};  smoothing  reduced  $\sigma  (\delta  v)$  by  half,  
but  the  statistical  properties  of  the  distribution  (described  below)  were  effectively  identical  for  the  two  analyses.  
Figure  \ref{fig:fig7}  shows  the  results  from  the  smoothed  analysis  of  \hcopnt .    As  in  previous  similar  figures,  the  top  panel  shows  a  scatterplot  with 
 $\sigma  (\delta  v)$ on  the  y-axis  plotted  against  $\delta  v$  on  the  x-axis.    Points  below  the  dashed  red  (solid  green)  line  differ  from  zero  
with  three  (five)  sigma  significance. The  lower  panel  shows  the  frequency  distribution  for  $\delta  v$.  The  mean  value  is  $-0.075  \pm  0.019$  and  the 
 median  value  is  $-0.075  \pm 0.017$. As  with  our  integrated  asymmetry  parameter $A$,  $\delta  v$  is  expected  to  be zero for  an  optically  thin  line.    
Figure  \ref{fig:fig8}  shows  the  distributions  in  $\delta  v$  and  its  error  for  \htcopnt.    The  mean ($-0.006 \pm0.010$) and  median  ($-0.006\pm0.009$) are  both consistent  with  zero,  as  expected.    

\begin{figure}
\begin{center}
\includegraphics[scale=1.0, angle=0]{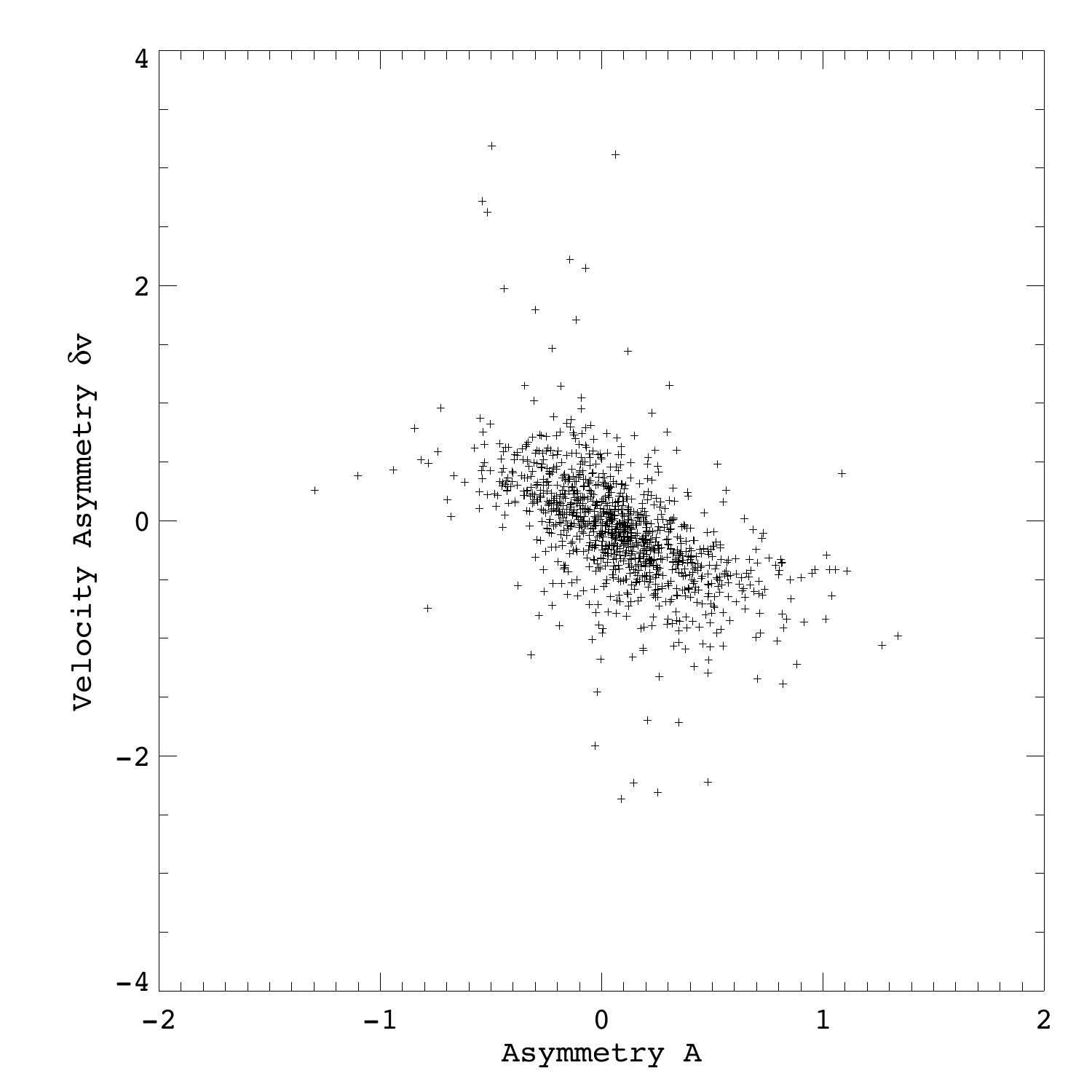} 
\caption{A comparison between the dervied values of the asymmetry parameter $A$ and the $\delta v$ parameter of \citet{Mardones1997}.  A clear anticorrelation is evident. }
\label{fig:fig9}
\end{center}
\end{figure}

Since  both  the  integrated  intensity  asymmetry  $A$  and  the  velocity  asymmetry  $\delta  v$  are  measures of a line's asymmetry,  we  would  expect  
that  they  are  closely related.  A distinct anti-correlation  is indeed  evident  in  the plot  of  $\delta  v$  against  $A$  for  \hcopnt\,  shown  in  Figure  \ref{fig:fig9}.   An anti-correlation is expected since positive values of $A$ but negative values of $\delta v$ indicate blue asymmetries, and vice versa for red asymmetries.  The Pearson correlation coefficient $R$ between $A$ and $\delta v$ is -0.516.
 
The  magnitude  of  the  negative  mean  of  $\delta  v$  for  the  MALT90  
sources, $-0.075\pm0.019$,  is somewhat  smaller  than  the  value  of  $-0.14  \pm  0.08$  that  \citet{Mardones1997}  observed  in  their  sample  of  47  low-mass  
star-forming  regions.
 Direct  comparison  of  the  two  samples  is  difficult  due  to the  disparate  selection criteria, resulting in two distinct samples, one low-mass and the other high-mass, and the different lines observed.    However,  it  is  interesting 
 to  note  that  the  FWHM  of  the  $\delta  v$  distribution  for  the  MALT90  sample  is  about  1,  roughly  the  same  as  the  width  of  the  $\delta  v$  distribution  in  Mardones  
(their  figure  3),  although  our  typical  \nthpnt\,    FWHM linewidth $\Delta V$  is  about  three  times  larger  than  in  their  sample.
 
\section{Discussion}

A summary of all of the asymmetry measurments using the three different parameters: the asymmetry parameter $A$, the local slope at the reference velocity, and $\delta v$ from \citet{Mardones1997} are presented in Table \ref{table:hcop_asym} for the optically thick \hcopnt\, line and Table \ref{table:htcop_asym} for the optically thin \htcopnt\, line.
For all three measures of the blue asymmetry,  the ensemble average for \hcopnt\, indicates a small but statistically significant blue asymmetry, but the ensemble average for \htcopnt\, shows no asymmetry within the errors.  The parameter with the most significant asymmetry measurement is the asymmetry parameter $A$, for which the measured value is positive by $> 8\sigma$ in the mean.  Furthermore, for all three parameters, the fraction of sources showing evidence for infall by having the predicted positive or negative sign is significantly greater than 50\% for \hcopnt\, and within the errors equal to 50\% for \htcopnt.

Thus, for the MALT90 sample of high-mass clumps, all three infall parameters for the optically thick \hcopnt\, line match the theoretical predictions of blue asymmetries caused by overall collapse motions of the star-forming clumps.  Moreover, all three infall parameters for the optically thin \htcopnt\, line are consistent within the errors of having no net asymmetry for the ensemble, a result also predicted by theory. In the presumably optically thicker sources where \htcopnt\, is detected and used as the reference line, the asymmetry is larger still, a result also predicted by the theory. This experimental verification of the theoretical predictions for an ensemble of clumps undergoing collapse strongly supports the collapse hypothesis.  It is difficult to imagine alternative explanations for the optically thick line to be on average blue asymmetric but the optically thin line to be on average exactly symmetric.   We conclude therefore that clumps are indeed undergoing global gravitational collapse.

\subsection{Asymmetries by Infrared Classification}
Having concluded that clumps are collapsing, we next investigate how the collapse motions  change with time as the clumps evolve.
To develop a schematic timeline to study the evolution of high-mass star-forming regions, the MALT90 team used {\it Spitzer} images (3 -- 24 $\mu$m) to classify clumps into broad categories \citep{Jackson2013, Guzman2015, Rathborne2016}.  The six categories are: `Quiescent' clumps (these are IR dark and hence cold and dense), `Protostellar' clumps (these have either extended 4.5 $\mu$m emission indicating shocked gas and/or a compact 24 $\mu$m point source indicating an embedded, accreting protostar), `CHII' and `H II regions' (these are IR bright in all bands and either compact or extended), and `PDRs' (photodissociation regions; these are indicated by bright PAH emission in the 8 \um\, band and represent the molecular/ionized gas interface at which the gas is excited by the UV radiation from a nearby high-mass star).  (It should be noted that the ``PDR'' category may not represent a distinct phase of clump evolution but may instead merely indicate the association or line-of-sight coincidence of a clump with a PDR interface. PDRs, however, only occur in the vicinity of high-mass stars with significant UV flux, so there is nevertheless some corresponce between this category and high-mass main sequence stars.)   We also used an additional category (referred to as `Unknown') to separate out clumps for which the IR signatures are either ambiguous or not obvious.  Since the dust emission from ATLASGAL traces material along the line of sight across the Galaxy but \textit{Spitzer} is more sensitive to nearby emission, many of these `uncertain' clumps may in fact lie at the far side  of the Galaxy.  

\cite{Guzman2015} analyzed {\it Herschel} and ATLASGAL data toward a large sample of clumps identified by ATLASGAL and included in the MALT90 survey to investigate how the clumps change as they evolve from category to category.  From the dust conintuum data, they derive  the dust temperature and the column density distribution for the clumps in these various IR categories and find strong evidence for clump evolution, with a monotonically increasing dust temperature as well as an initial increase, followed by a subsequent decrease, of the column density with evolutionary stage.  Here we investigate whether the asymmetry parameter $A$ also varies with the evolutionary stage.

\begin{figure}
\begin{center}
\includegraphics[scale=1.0, angle=0]{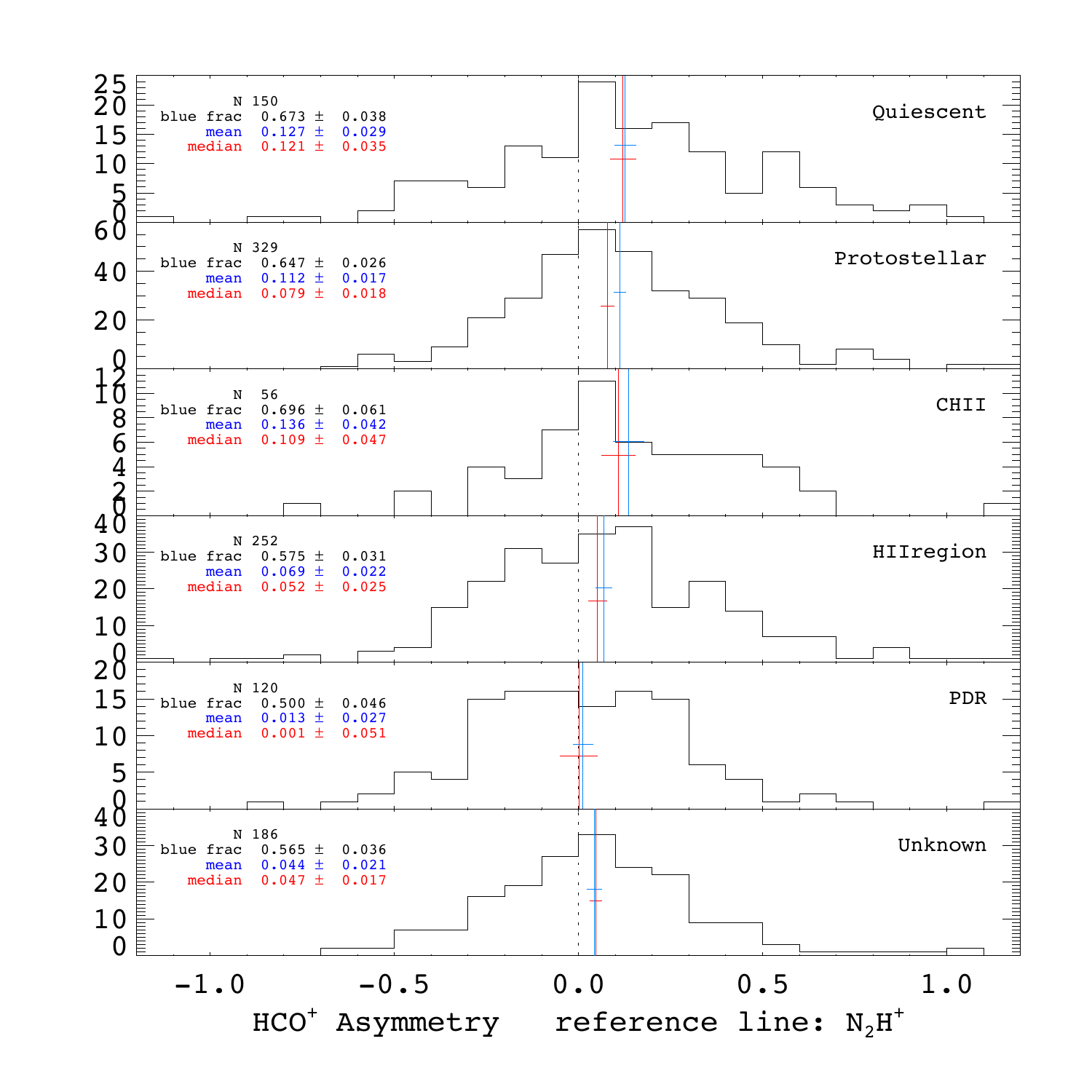} 
\caption{A histogram of all measured \hcopnt\, asymmetries for each evolutionary IR classification based on {\it Spitzer} images.The vertical red and blue lines and text denote the median and mean of the distribution; the horizontal red and blue lines indicate the $\pm 1 \sigma$ uncertainties in the median and mean. }
\label{fig:fig10}
\end{center}
\end{figure}

\begin{figure}
\begin{center}
\includegraphics[scale=1.0, angle=0]{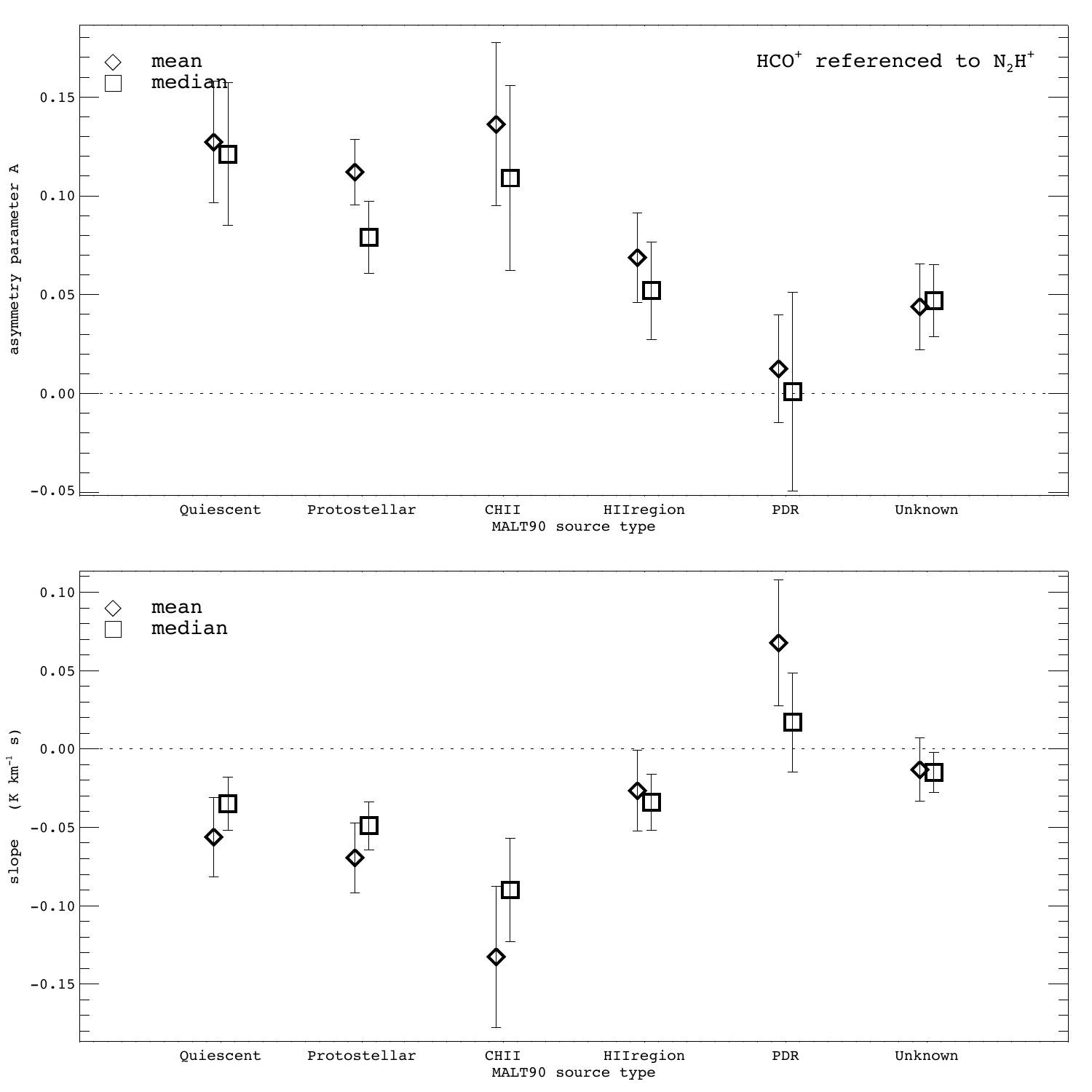} 
\caption{The mean and median \hcopnt\, asymmetries and line slopes plotted versus the IR classification. }
\label{fig:fig11}
\end{center}
\end{figure}

\begin{figure}
\begin{center}
\includegraphics[scale=1.0, angle=0]{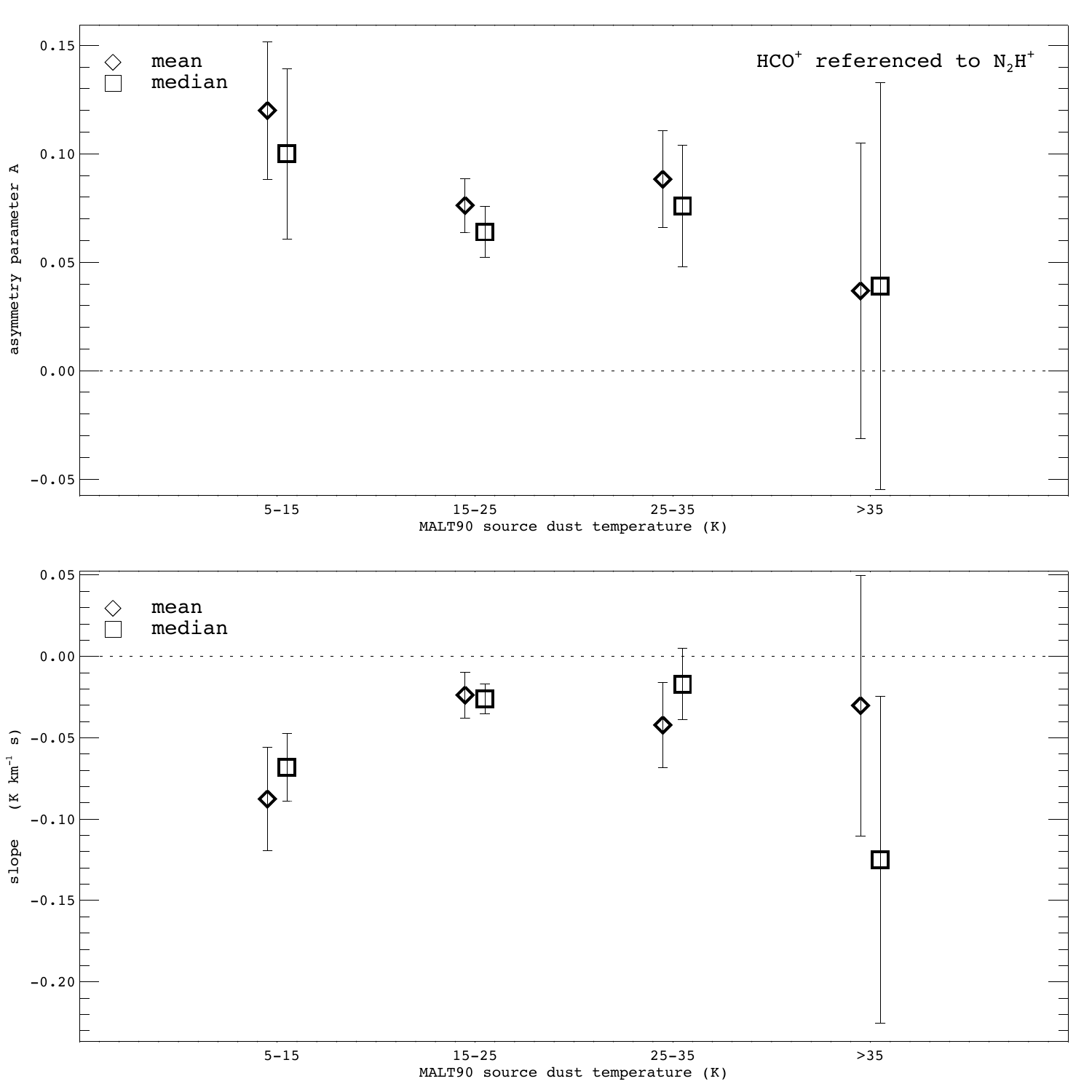} 
\caption{The mean and median \hcopnt\, asymmetries and line slopes plotted versus the source dust temperature. }
\label{fig:fig12}
\end{center}
\end{figure}

Figure \ref{fig:fig10} shows the \hcopnt\, asymmetry distributions for each of our IR classes.  This plot shows significantly positive values of $A$ for clumps at earlier stages of their evolution, and no collapse for clumps in the PDR category.  The PDR asymmetries shows a flatter distribution than is seen for the other categories; a K-S test asking if this distribution could be drawn from the summed distribution of the four classes above it in Figure \ref{fig:fig10} gave a probability of $\sim$3\%.    Figure \ref{fig:fig11} shows the mean and median asymmetries and slopes plotted against our IR classifications and in Figure \ref{fig:fig12} against dust temperature.   There is no strong variation of $A$ with dust temperature, although there is a marginal indication of a slight decrease in $A$ with dust temperature.  As shown in \citet{Guzman2015} the temperature ranges of the different IR classes overlap considerably. Thus, the infrared classification may better indicate the clumps' evolutionary stage than the {\it Herschel} dust temperature.

The fact that the asymmetry parameter $A$ is largest for the quiescent, protostellar, and compact \hii\, region categories, smaller for the \hii\, region category, and near 0 for the PDR category
suggests that clump collapse is already well underway in the earliest quiescent phase, continues through the protostellar phase, subsides in the \hii\, region phases, and ends by the time a PDR is formed.     Alternatively, clump  collapse may continue through these later phases, but the blue asymmetry collapse signal may well be hidden by expansion motions of the \hii\, regions.  

At first glance it is somewhat surprising that quiescent clumps show the blue asymmetry.  These quiescent clumps presumably represent an evolutionary stage before high-mass stars have formed within them.  Thus they should lack internal heating sources  usually invoked to explain the blue asymmetry.  If the inner regions of the clump have the same excitation temperature as the outer regions, no asymmetry can occur even in a collapsing clump.  One possible explanation for the clear presence of the blue asymmetry collapse signature in quiescent clumps is that the clumps have a significant density gradient, with the inner regions denser than the outer regions.  If the densities in the inner regions of the clump exceed the critical density of the \hcop\, line, but the densities in the outer regions are smaller than the critical density, then the thermalized inner regions will indeed have a larger excitation temperature than the subthermally excited outer regions.  In this case, the blue asymmetry would still occur for a collapsing clump even in the absence of internal heating sources.  Alternatively, it may also be possible that low-mass star formation is occurring in the inner regions of quiescent clumps, and these low mass stars act as an internal heating source. If this is the case, then the low-mass stars  must be insufficiently luminous to be detected due to the clumps' very large extinction.  In either case, our detection of collapse signatures in the quiescent clumps agrees with the study of \cite{Contreras2018}, who find a clear collapse signature toward a high-mass core in a pre-stellar, 70 \um\, dark clump.

\section{Summary}
\label{sec:summary}
Using molecular line data from the MALT90 Survey toward a large sample of 1,093 clumps, we have searched for the blue asymmetry, a spectral indicator of collapse, using the \hcop\,  line as an optically thick collapse tracer and the \nthp\, line as an optically thin systemic velocity tracer.  In order to avoid complications arising from non-Gaussian line shapes, we use an asymmetry parameter $A$ defined as the fraction of the total line flux lying blueward of the systemic velocity.  With this definition positive values of $A$ indicate collapse motions, and negative values expansion motions.
From this analysis the following conclusions are reached:

1. The ensemble of clumps shows statistical evidence for gravitational collapse.  Both the mean ($0.083 \pm 0.010$) and median ($ 0.065 \pm 0.009$) for the distribution of the measured values of $A$ are positive and statistically significant, indicating a preponderance of collapse motions in the clump sample.  Moreover, more clumps ($0.607 \pm 0.015$ of the entire sample) show positive values of $A$ than negative values.  Since theoretical investigations of simulated clumps suggest that the collapse signature can be easily obscured, this analysis suggests that the dense molecular clumps identified by ATLASGAL are in general undergoing gravitational collapse.

2. The result was checked using the optically thin \htcop\, line.  As expected, the asymmetry parameter using \htcop\, as the collapse indicator and \nthpnt\, as the reference line was close to zero.

3. When \hcopnt\, was used as the collapse tracer but \htcopnt\, was used instead of \nthpnt\, as a reference line, the value of $A$ was slightly larger, consistent with the larger expected asymmetries for sources with higher \hcopnt\, optical depths.  However, since \nthpnt\, is more readily detected than \htcopnt, it can be used to characterize the collapse toward a larger sample and with better determined values of $A$.

4. Two alternate methods to measure blue asymmetries: (1) the slope of the optically thick line at the reference velocity and (2) the $\delta v$ parameter introduced by \cite{Mardones1997}, also indicate statistically significant blue asymmetries for \hcopnt\, but no asymmetry for \htcopnt, albeit with less statistical significance than that for $A$.  The fact that three independent methods detect the blue asymmetry for the optically thick \hcopnt\, line but no asymmetry for the optically thin \htcopnt\, line provides strong evidence that the ensemble of clumps is undergoing gravitational collapse.

5.  Using {\it Spitzer} infrared images to estimate the evolutionary stage of the clumps, we found that the earliest stages have larger positive values of $A$ than the \hii\, region stage.  The PDR stage shows no evidence for global collapse.  This result indicates that collapse is already underway in the earliest quiescent stage but the collapse signature diminishes as the \hii\, region expands.

6.  The presence of a blue asymmetry in the earliest quiescent stage, before high-mass stars have formed, suggests either a large gradient in density or internal heating from undetected low-mass stars.

\section{Acknowledgements} 
At the time of data collection, the Mopra radio telescope used for the MALT90 survey was operated by the Australia Telescope National Facility which is funded by the Australian Government for operation as a National Facility managed by CSIRO.  J.M.J. gratefully acknowledges funding support from the US National Science Foundation Grant No. AST-0808001 and the CSIRO Distinguished Visitor program.  PS was financially supported by Grant-in-Aid for Scientific Research (KAKENHI Number 18H01259) of
Japan Society for the Promotion of Science (JSPS).  The MALT90 team is grateful for the help and support of our colleagues in the ATLASGAL survey team, especially Frederic Schuller and Frederick Wyrowski, who kindly provided the ATLASGAL positions and fluxes of survey sources ahead of publication.  These ATLASGAL survey sources are the target list for the MALT90 Survey.  We thank Francesca Schiavelli for assistance in an early analysis of the asymmetric line profiles. This research was conducted in part at the SOFIA Science Center, which is operated by the Universities Space Research Association under contract NNA17BF53C with the National Aeronautics and Space Administration.

\section{Appendix}
\subsection{Rest frequencies and velocity corrections}
The asymmetry analysis relies upon  accurate values for the rest frequencies of the primary and reference lines.  A shift of 10 kHz in a 90 GHz rest frequency leads to a velocity shift of 0.033 \kms, which will cause a shift in the asymmetry parameter $A$ that is larger than the typical error in the mean or median of the distribution.  In the course of this work we discovered some small discrepancies between the rest frequencies given in the MALT90 data file headers, the frequencies actually used in the data processing pipeline, and the current best values of these frequencies.  Using those current best values we calculated velocity corrections $\delta V$ to be subtracted from the MALT90 catalog velocities \citep{Rathborne2016}.  Table \ref{table:freqs} lists these corrections.  Uncertainties in the rest frequencies will result in systematic errors in any velocity calculated from the difference between an observed frequency and that rest frequency; the standard errors in the rest frequencies $\sigma_f$  and the corresponding velocity uncertainties $\sigma_V$ are listed in the last two columns of the table.  The data for \hcopnt\, were taken from the Splatalogue (http://www.cv.nrao.edu/php/splat/).  The updated  \htcopnt\, frequency is the optically thin intensity-weighted mean frequency (i.e., weighted according to the statistical weights) for the unresolved 0.133 \kms\, doublet splitting reported in \citet{Schmid-Burgk2004}.  The updated \nthpnt\, frequency is the optically thin intensity-weighted average of the frequencies of the unresolved $JF_{1}F $ 12x---01x transitions, from \citet{Pagani2009}.  The hyperfine spacings found by \citet{Pagani2009} were in excellent agreement with the spacings determined by \citet{Caselli1995}; we used the frequencies of \citet{Pagani2009}  as their absolute frequency uncertainty was 0.004 MHz while the corresponding value for  \citet{Caselli1995}  was 0.007 MHz.  

\subsection{Fitting the \nthpnt spectra}
\begin{figure}
\begin{center}
\includegraphics[scale=0.751, angle=0]{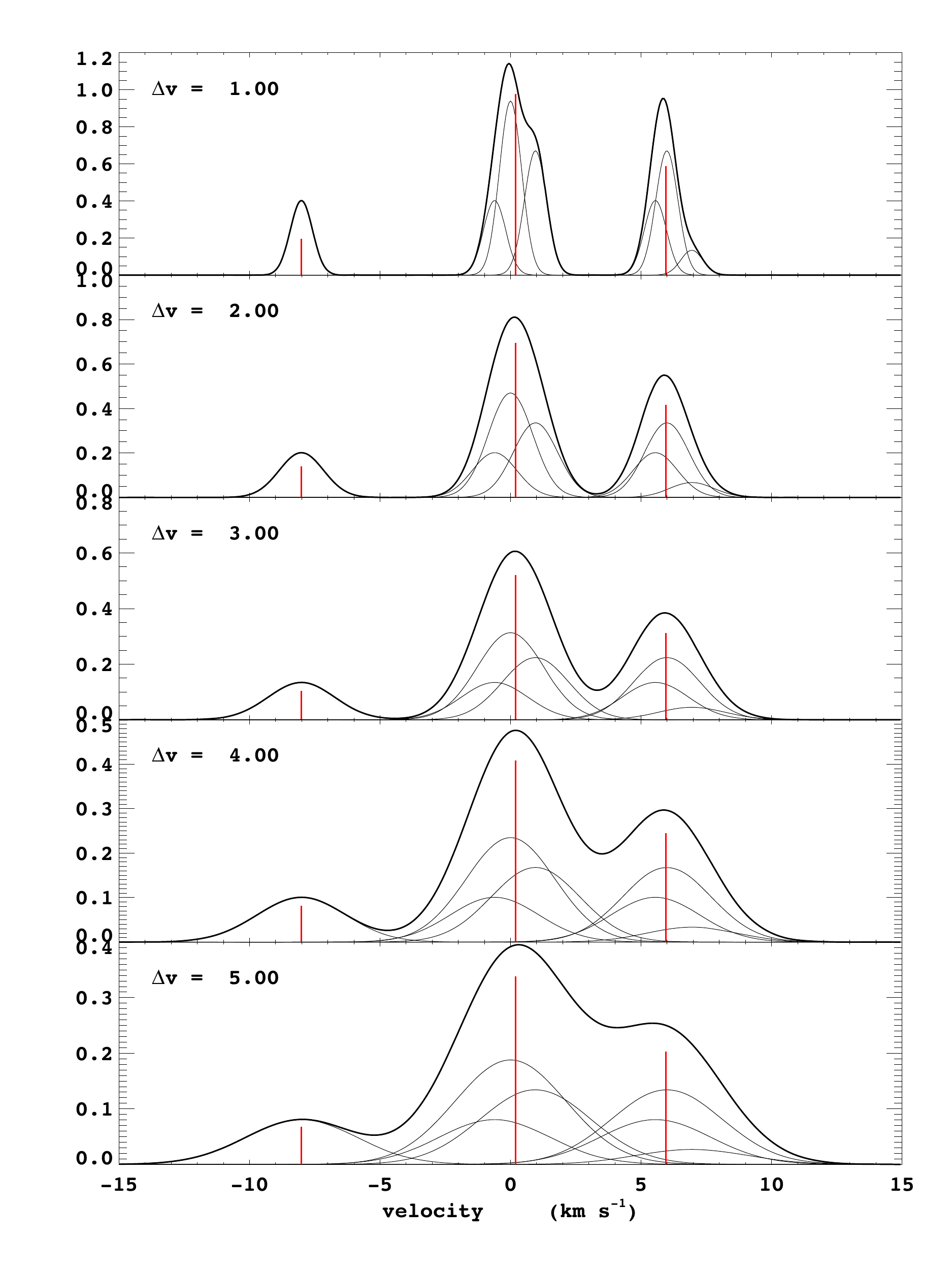} 
\caption{Synthesized spectra of \nthpnt\, at different velocity widths.  See the text in the Appendix for explanation of the various curves. }
\label{fig:fig13}
\end{center}
\end{figure}

The 93 GHz \nthp\, spectrum is the sum of fifteen different transitions at seven different frequencies.  For velocity widths (FWHM) greater than $\sim$1 \kms\, the hyperfine splitting due to the nuclear spin of the inner nitrogen atom (the F quantum number) is unresolved and  lines merge into three components: a central  triplet ($JF_1=12-01$) with an intensity-weighted frequency of 93.173703 GHz, offset by +0.196 \kms\, from the most intense individual hyperfine line $JF_1F=123-012$; a triplet ($JF_1=11-01$) with weighted frequency corresponding to +5.747 \kms\, from the intensity-weighted position of the main component; and a single line ($JF_1=10-01$) at -8.203 \kms\, from the main component.  As the velocity width increases the line profile will approach smooth Gaussian-like line shapes, as illustrated in Figure \ref{fig:fig13}.   We assume  all fifteen transitions have a common velocity width $\Delta V$.  We further assume that there all lines are optically thin and that the lines have a common excitation temperature.  In  Figure \ref{fig:fig13} the thin black curves indicate the seven lines, and the thick black curve is the summed line profile.  The red vertical lines indicate the grouped line strengths at the intensity-weighted line positions, in the usual 1:5:3 relative normalization.  

\begin{figure}
\begin{center}
\includegraphics[scale=1.0, angle=0]{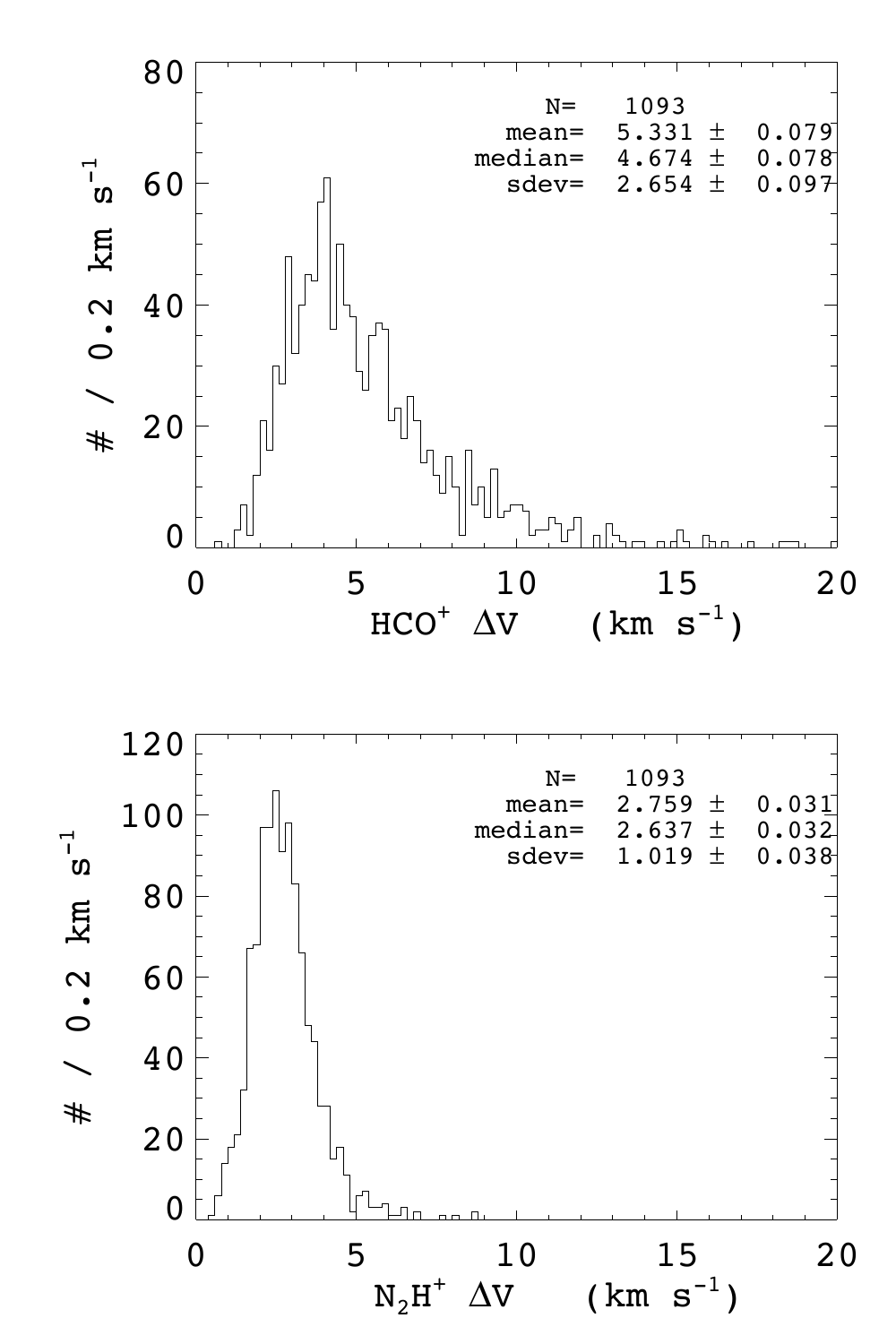} 
\caption{FWHM velocity widths for \hcopnt\, (upper panel) and \nthpnt\, (lower panel) for the MALT90 sources included in the \hcopnt\, asymmetry analysis.}
\label{fig:fig14}
\end{center}
\end{figure}

Fitting this complex spectrum was made more challenging by the modest signal/noise of the MALT90 dataset: typical noise was 0.18 K while amplitudes were $\sim$1 K.  Further, we observed relatively large line widths for most of our sources; Figure \ref{fig:fig14} shows the fitted FWHM line widths for \hcopnt\, and \nthpnt\, for the sources included in the \hcopnt\, asymmetry analysis.  It is evident from Figure \ref{fig:fig13} that a fit with seven independent components would have very poor convergence properties due to the high degree of overlap.  In our previous work \citep{Rathborne2016} we fit the \nthp\, spectrum with three independent Gaussians with a common width and the centers of the hyperfine satellites  constrained to the established velocity differences.  Studies with synthetic spectra where we simulated all seven lines with a common velocity width, superposed on a signal-free MALT90 noise spectrum, indicated this model slightly over-estimated the velocity width, that there was a systematic offset in the fit velocity of -0.01 to -0.02 \kms , and that the fit error in the systemic velocity was underestimated by $\sim$15\%.  Consequently, for the present analysis we used a more complex model with seven Gaussian profiles, all with a common width parameter: the left component constituted a single Gaussian, while the center and right components were each represented as the superposition of three Gaussian profiles with spacing constrained to the \citet{Pagani2009} values and relative amplitudes within each group of three fixed according to their statistical weights.  The positions of the left and right components were fixed at the established spacing from the intensity-weighted mean position of the center component.  Thus, the fitted parameters were the velocity of the center component, the velocity width of the seven constituent Gaussian profiles, and three amplitudes for the left, center, and right components.  Specifically, the fit function for intensity as a function of velocity $v$ was
\begin{multline*}
I(v) =  A_L(1.000e^{-(v+8.204-V)^2/2S^2} )\\
+ A_C(0.200e^{-(v+0.810-V)^2/2S^2} + 0.466e^{-(v+0.196-V)^2/2S^2} + 0.334e^{-(v-0.760-V)^2/2S^2}) \\
+ A_R(0.333e^{-(v-5.351-V)^2/2S^2} + 0.556e^{-(v-5.786-V)^2/2S^2} + 0.111e^{-(v-6.738-V)^2/2S^2})
\end{multline*}
where the fitted variables are the systemic velocity V, the common velocity dispersion S, and the amplitudes $A_L$,  $A_C$, and $A_R$  for the left, center, and right components. The numerical coefficients of the Gaussians sum to unity for each component and are proportional to the statistical weights of the lines; and the velocity offsets in the exponents are taken from \citet{Pagani2009}, relative to the optically thin intensity-weighted average for the center component.  This rather \textit{ad hoc} model was motivated by the limited Signal/Noise and the large observed velocity widths  of the MALT90 spectra.  Under the assumptions of negligible optical depth and uniform excitation temperatures for all transitions, the three amplitudes $A_L$,  $A_C$, and $A_R$ would be expected to be in the ratio 1:5:3 according to the statistical weights; this is what was was generally observed, but with large errors in consequence of the limited S/N of the data.

Studies with synthetic spectra showed this model eliminated the systematic velocity offset and provided an accurate estimate of the underlying velocity width.  The under-estimation of the velocity error persisted; consequently we increased the errors of the fit to the \nthpnt\, systemic velocity by a factor of 1.15 when calculating the asymmetry errors.  We undertook several studies to test the validity of this model for our data.

Non-negligible \nthpnt\, optical depths would modify the line shapes illustrated in Figure \ref{fig:fig13} and would shift the fitted velocities.  In the limit of high optical depth (while maintaining uniform excitation temperatures), the observed intensities of the three contributors to the main component would be equal and the weighted position would shift by -0.082 \kms, while the ``left'' lowest-velocity hyperfine component, being a single transition, would not be affected by large optical depth.   To test our three-component/seven-Gaussian model and to look for evidence of optically thick \nthpnt\, emission, we reanalyzed the \nthpnt\, spectra of the 1093 sources contributing to our \hcopnt asymmetry analysis.  We fit a single Gaussian to the spectrum in the range $\pm 5$ \kms\, about the position of the left hyperfine peak as expected on the basis of the model fit.  This is the lowest intensity component of a spectrum that is already S/N-challenged.  To select reasonably good fits, we required that the integrated intensity S/N of the single-Gaussian fit be $>$4 and that the velocity width for the single-Gaussian fit and the model fit be $<$ 4 \kms.   

The results of this study are shown in Figure \ref{fig:fig14}.   Of the 1093 sources in the sample, 275 passed the selection cuts.   A Gaussian fit to the velocity offset distribution in the left panel has a mean of -0.020 \kms\, and a dispersion of 0.164 \kms\, (which is dominated by the uncertainty of the single-gaussian fit velocity, which had a median value of 0.13 \kms).  The uncertainty in the mean velocity offset is $\sigma_{mean}=0.164/\sqrt{275}=0.010$ \kms.  This offset of $-0.020 \pm 0.010$ \kms\, is small compared to the predicted optically thick offset of -0.082 \kms\,  noted above, supporting our assumption that \nthpnt\, was generally optically thin.  We conclude that our model fit  provides robust velocity estimates with a systematic error on the order of 0.02 \kms.  

\begin{figure}
\begin{center}
\includegraphics[scale=1.0, angle=0]{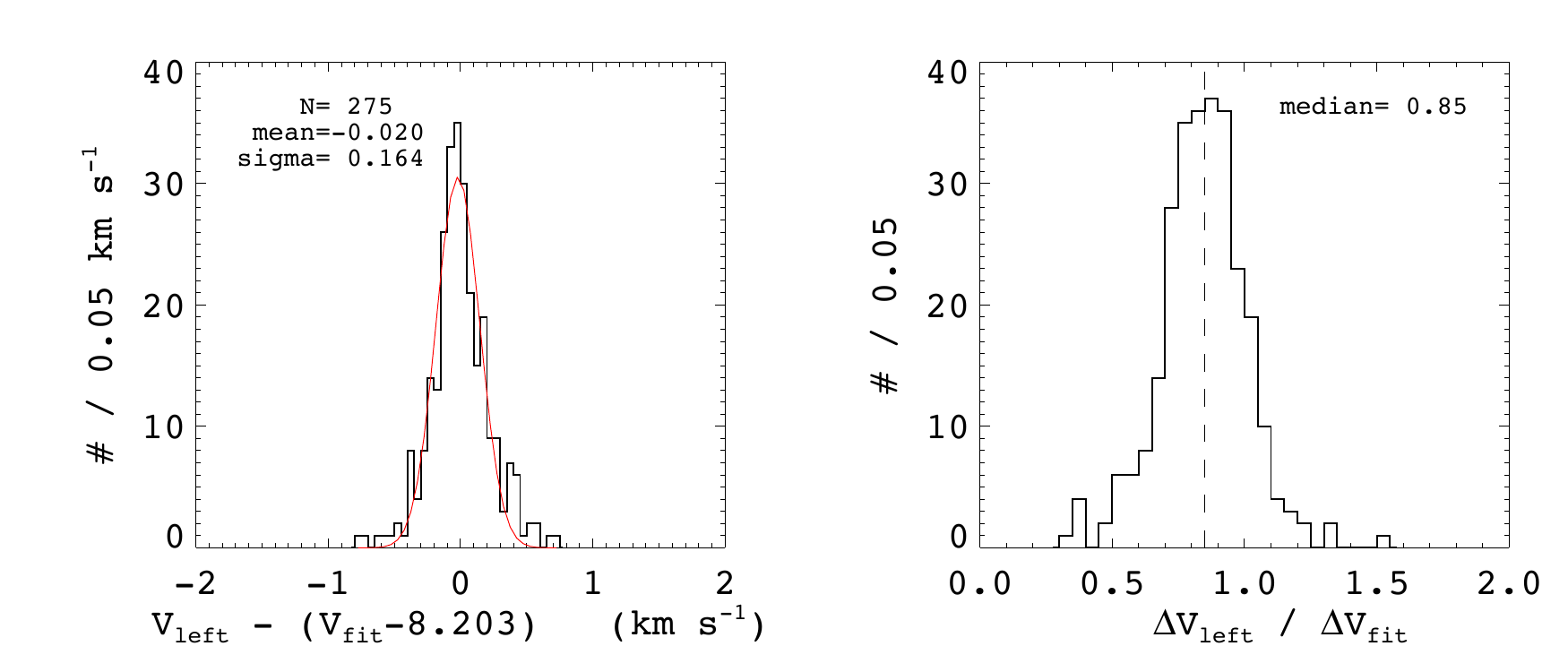} 
\caption{Results from fitting single Gaussians to the left \nthpnt\, hyperfine component.    Left:  histogram of the difference between the single-Gaussian fit velocity and the nominal velocity offset from the fit velocity, with a Gaussian fit to the velocity difference shown in red.  Right: the ratio of the single-Gaussian velocity width to the width returned in the model fit. }
\label{fig:fig15}
\end{center}
\end{figure}

The right panel of Figure \ref{fig:fig15} shows the distribution in the ratio of the width of the left component to the common width returned by our model fit.  The median of this distribution is 0.85 rather than the expected  1.0  that we observed in our synthetic spectra studies.  This could indicate the presence of optically thick emission, self-absorption, or of intensity anomalies due to non-LTE excitation of the seven constituent transitions.  Such effects would likely tend to equalize the amplitudes of the three constituent lines in the center or right component, broadening the apparent velocity width.  Intensity anomalies have often been detected in the \nthp\, observations \citep[see e.g.][]{Caselli1995, Daniel2006, Daniel2007}. The large widths and limited signal/noise of our data set preclude definitive conclusions.   Non-negligible \nthpnt\, optical depth would shift the \nthpnt\, velocity slightly negatively, which would reduce the Blue integration and increase the Red integration of the primary spectrum --  decreasing the observed Blue/Red asymmetry and shifting $A$ to more negative values.   Hence, untreated optical depth effects cannot be responsible for the positive Blue/Red asymmetries we observe in our data.  

\begin{figure}
\begin{center}
\includegraphics[scale=1.0, angle=0]{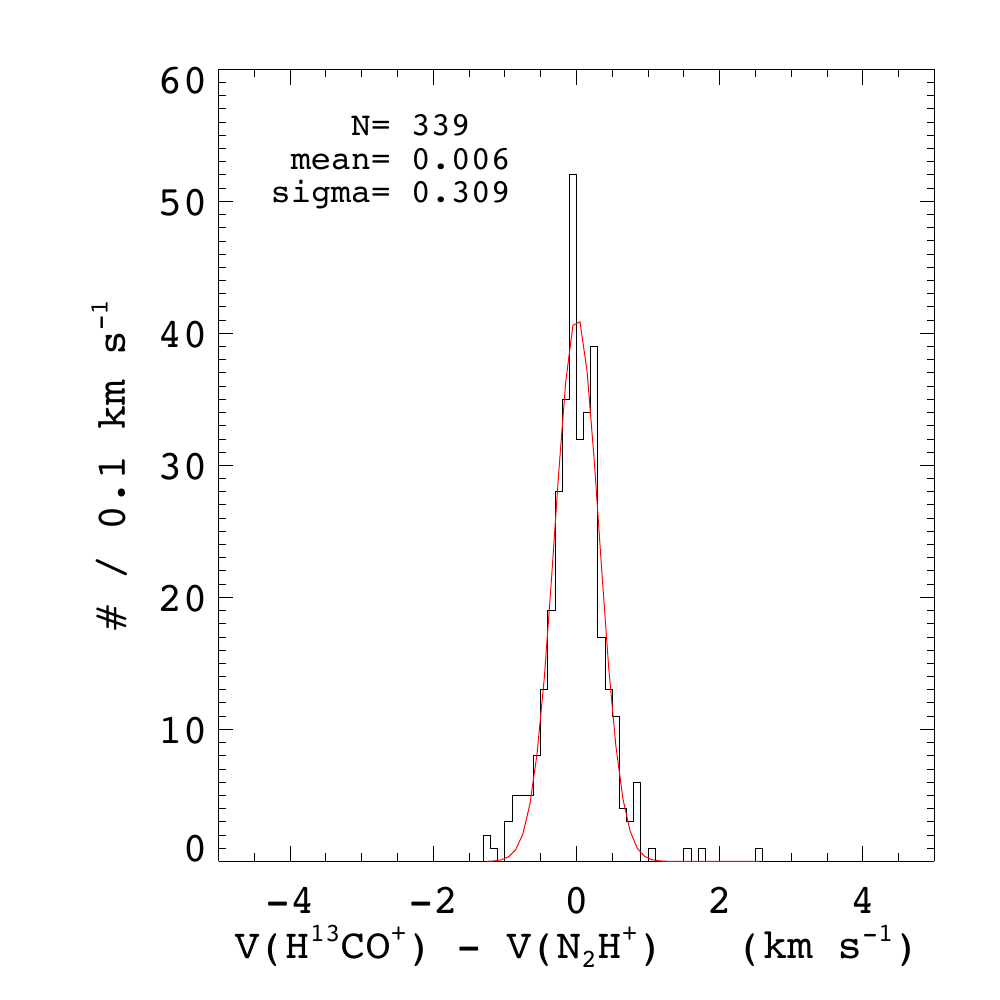} 
\caption{Difference between the fitted velocities for \nthpnt\, and \htcopnt, with a Gaussian fit to the velocity difference shown in red.}
\label{fig:fig16}
\end{center}
\end{figure}

We gain further confidence in the precision of our fitted \nthpnt\, velocities from a comparison with the fitted velocities of \htcopnt\, for the 339 sources where both lines are detected.  Both lines are expected to be optically thin, and it is reasonable to expect their velocities to be in good agreement.  The distribution in velocity differences is shown in Figure \ref{fig:fig16}; a Gaussian fit yields a mean velocity difference of 0.006 \kms, with a $1\sigma$ dispersion of 0.309 \kms\, and a corresponding uncertainty in the mean difference of 0.017 \kms.  The systematic uncertainty in the velocity difference due to the frequency uncertainties listed in Table \ref{table:freqs} is 0.021 \kms, in excellent agreement with the data.  Using the \citet{Caselli1995} rest frequency for \nthpnt\, results in a mean velocity difference with \htcopnt\, of -0.044 \kms; this differs from our expectation of a mean difference of zero by 2.6 times the statistical uncertainty but just twice the systematic error.  Thus, while these results favor the \citet{Pagani2009} rest frequency, our statistics do not allow a definitive finding.

In consequence to these studies, we adopted $\pm 0.03$ \kms\, as the estimate of systematic uncertainty in our \nthpnt\, reference velocity, and we varied the reference velocities by this amount to estimate the systematic uncertainties in the means and medians of our observed asymmetry distributions.

\newpage
\begin{deluxetable}{cccc}
\tablecaption{Estimated \hcopnt\, Systematic Asymmetry Parameter Errors \label{table:syserrs}}
\tablehead{\colhead{\nthpnt\, Velocity} & \colhead{Mean Asymm.} &  \colhead{Median Asymm.} & \colhead{Blue Fraction} }
\startdata
nominal  & 0.083 $\pm$ 0.010 & 0.065 $\pm$ 0.009 & 0.607$\pm$ 0.015 \\
$+ 0.03$ \kms & 0.094 $\pm$ 0.010 & 0.083 $\pm$ 0.010  & 0.624 $\pm$ 0.015\\
$- 0.03$ \kms& 0.072 $\pm$ 0.010 & 0.057 $\pm$ 0.008 & 0.590 $\pm$ 0.015\\
\enddata
\tablecomments{Variation of the mean of the asymmetry parameter $A$, the median of the  asymmetry parameter $A$, and the ``Blue Fraction,'' the fraction of the sample with positive values of $A$ for the \hcop\, line, using  \nthp\, to determine the reference velocity, as the reference velocity is offset by its estimated systematic uncertainty of 0.03 \kms.}
\end{deluxetable}

\begin{deluxetable}{crrr}
\tablecaption{\hcopnt Asymmetry Using \nthpnt as the Reference \label{table:hcop_asym}}

\tablehead{\colhead{Parameter} & \colhead{Mean} &  \colhead{Median} & \colhead{Blue Fraction} }
\startdata
$A$ & $0.083 \pm 0.010$  & $0.065 \pm 0.009$ & $0.607 \pm 0.015$  \\ 
Slope at $V_{ref}$ & $-0.036\pm 0.012$  & $-0.034 \pm 0.009$ & $0.562 \pm 0.015$  \\
$\delta v$ & $-0.075 \pm 0.019$  & $-0.075 \pm 0.017$ & $0.578 \pm 0.015$   \\
\enddata
\tablecomments{The expected sign for a ``blue'' asymmetry is positive for $A$ and negative for the Slope at $V_{ref}$ and for $\delta v$.  
The slope is $d(T_A^*)/d(V_{LSR})$
 in units of K km$^{-1}$ s. 
The ``Blue Fraction'' indicates the fraction of sources having the positive or negative sign for the parameter consistent with a ``blue'' asymmetry.}
\end{deluxetable}

\begin{deluxetable}{crrr}
\tablecaption{\htcopnt Asymmetry Using \nthpnt as the Reference \label{table:htcop_asym}}

\tablehead{\colhead{Parameter} & \colhead{Mean} &  \colhead{Median} & \colhead{Blue Fraction} }
\startdata
$A$ & $0.000 \pm 0.014$  & $-0.015 \pm 0.020$ & $0.487 \pm 0.027$  \\ 
Slope at $V_{ref}$ & $-0.002 \pm 0.007$  & $0.000 \pm 0.008$ & $0.499 \pm 0.007$  \\
$\delta v$ & $-0.006 \pm 0.010$  & $-0.006 \pm 0.009$ & $0.522 \pm 0.027$   \\
\enddata
\tablecomments{The expected sign for a ``blue'' asymmetry is positive for $A$ and negative for the Slope at $V_{ref}$ and for $\delta v$. 
The slope is $d(T_A^*)/d(V_{LSR})$
 in units of K km$^{-1}$ s. 
 The ``Blue Fraction'' indicates the fraction of sources having the positive or negative sign for the parameter consistent with a ``blue'' asymmetry.}
\end{deluxetable}

\begin{deluxetable}{ccccccc}


\tablecaption{Frequency corrections and velocity offsets \label{table:freqs}}

\tablehead{\colhead{Line} & \colhead{header freq} &  \colhead{pipeline freq} & \colhead{updated freq} & \colhead{$\delta V$} & \colhead{$\sigma_f$} & \colhead{$\sigma_{V}$} \\
\colhead{(---)} & \colhead{(MHz)} & \colhead{(MHz)} & \colhead{(MHz)} & \colhead{(km/s)} & \colhead{(MHz)} & \colhead{(km/s)} }
\startdata
\hcop &89188.523 &89188.526 &89188.525 &0.004 &0.004 &0.013\\
\nthp	&93173.711 &93173.711 &93173.703 &0.027 &0.004&0.013 \\
\htcop &86754.328 &86754.330 &86754.288 &0.144 &0.005 &0.017\\
\enddata

\tablecomments{ $\delta V$ is the velocity correction derived from the difference between the pipeline and the updated frequencies; it is to be subtracted from the MALT90 catalog velocity values \citep{Rathborne2016}.  $\sigma_f$ is the uncertainty in the frequency, and $\sigma_{V}$ is the corresponding systematic uncertainty in any velocity derived using that frequency.}
\end{deluxetable}

\newpage

\newpage
\begin{thebibliography}{dummy}

\bibitem[Bonnell et al.(1997)]{Bonnell1997} Bonnell, I.~A., Bate, M.~R., Clarke, C.~J., \& Pringle, J.~E.\ 1997, \mnras, 285, 201
\bibitem[Campbell et al.(2016)]{Campbell2016} Campbell, J.~L., Friesen, R.~K., Martin, P.~G., et al.\ 2016, \apj, 819, 143 
\bibitem[Caselli et al.(1995)]{Caselli1995} Caselli, P., Myers, P.~C., \& Thaddeus, P.\ 1995, \apjl, 455, L77		
\bibitem[Chen et al.(2010)]{Chen2010} Chen, X., Shen, Z.-Q., Li, J.-J., Xu, Y., \& He, J.-H.\ 2010, \apj, 710, 150 		
\bibitem[Chira et al.(2014)]{Chira2014} Chira, R.-A., Smith, R.~J., Klessen, R.~S., Stutz, A.~M., \& Shetty, R.\ 2014, \mnras, 444, 874 	
\bibitem[Contreras et al.(2013)]{Contreras2013}Contreras, Y., Schuller, F., Urquhart, J.~S., et al.\ 2013, \aap, 549, A45
\bibitem[Contreras et al.(2017)]{Contreras2017} Contreras, Y., Rathborne, J.~M., Guzman, A., et al.\ 2017, \mnras, 466, 340
\bibitem[Contreras et al.(2018)]{Contreras2018} Contreras, Y., Sanhueza, P., Jackson, J.M.., et al.\ 2018, \apj, submitted
\bibitem[Daniel et al.(2006)]{Daniel2006} Daniel, F., Cernicharo, J., \& Dubernet, M.-L.\ 2006, \apj, 648, 461
\bibitem[Daniel et al.(2007)]{Daniel2007} Daniel, F., Cernicharo, J., Roueff, E., Gerin, M., \& Dubernet, M.~L.\ 2007, \apj, 667, 980	
\bibitem[Evans(2003)]{Evans2003} Evans, N., II 2003, SFChem 2002: Chemistry as a Diagnostic of Star Formation, 157 
\bibitem[Foster et al.(2011)]{Foster2011}Foster, J.~B., Jackson, J.~M., Barnes, P.~J., et al.\ 2011, \apjs, 197, 25
\bibitem[Foster et al.(2012)]{Foster2012} Foster, J.~B., Stead, J.~J., Benjamin, R.~A., Hoare, M.~G., \& Jackson, J.~M.\ 2012, \apj, 751, 157
\bibitem[Foster et al.(2011)]{Foster2011} Foster, J.~B., Jackson, J.~M., Barnes, P.~J., et al.\ 2011, \apjs, 197, 25
\bibitem[Foster et al.(2013)]{Foster2013} Foster, J.~B., Rathborne, J.~M., Sanhueza, P., et al.\ 2013, \pasa, 30, e038		
\bibitem[Fuller et al.(2005)]{Fuller2005} Fuller, G.~A., Williams, S.~J., \& Sridharan, T.~K.\ 2005, \aap, 442, 949 		
\bibitem[Gregersen et al.(1997)]{Gregersen1997} Gregersen, E.~M., Evans, N.~J., II, Zhou, S., \& Choi, M.\ 1997, \apj, 484, 256 
\bibitem[Guzm{\'a}n et al.(2015)]{Guzman2015} Guzm{\'a}n, A.~E., Sanhueza, P., Contreras, Y., et al.\ 2015, \apj, 815, 130		
\bibitem[Ho \& Haschick(1986)]{HoHaschick1986} Ho, P.~T.~P., \& Haschick, A.~D.\ 1986, \apj, 304, 501 	
\bibitem[Hoq et al.(2013)]{Hoq2013} Hoq, S., Jackson, J.~M., Foster, J.~B., et al.\ 2013, \apj, 777, 157 	
\bibitem[Jackson et al.(2013)]{Jackson2013}Jackson, J.~M., Rathborne, J.~M., Foster, J.~B., et al.\ 2013, \pasa, 30, e057
\bibitem[Keto et al.(1987)]{Keto1987} Keto, E.~R., Ho, P.~T.~P., \& Reid, M.~J.\ 1987, \apjl, 323, L117	
\bibitem[Krumholz et al.(2005)]{KrumholzMcKeeKlein2005} Krumholz, M.~R., McKee, C.~F., \& Klein, R.~I.\ 2005, \nat, 438, 332 	
\bibitem[Lee et al.(1999)]{Lee1999} Lee, C.~W., Myers, P.~C., \& Tafalla, M.\ 1999, \apj, 526, 788 		
\bibitem[Leung \& Brown(1977)]{LeungBrown1977} Leung, C.~M., \& Brown, R.~L.\ 1977, \apjl, 214, L73 		
\bibitem[L{\'o}pez-Sepulcre et al.(2010)]{Lopez2010} L{\'o}pez-Sepulcre, A., Cesaroni, R., \& Walmsley, C.~M.\ 2010, \aap, 517, A66 		
\bibitem[Mardones et al.(1997)]{Mardones1997} Mardones, D., Myers, P.~C., Tafalla, M., et al.\ 1997, \apj, 489, 719 
\bibitem[McKee \& Ostriker(2007)]{McKeeOstriker2007} McKee, C.~F., \& Ostriker, E.~C.\ 2007, \araa, 45, 565	
\bibitem[Myers et al.(1996)]{Myers1996} Myers, P.~C., Mardones, D., Tafalla, M., Williams, J.~P., \& Wilner, D.~J.\ 1996, \apjl, 465, L133 	
\bibitem[Myers(2005)]{Myers2005} Myers, P.~C.\ 2005, \apj, 623, 280	
\bibitem[Motte et al.(2017)]{Motte2017} Motte, F., Bontemps, S., \& Louvet, F.\ 2017, arXiv:1706.00118 	
\bibitem[Narayanan et al.(1998)]{Narayanan1998} Narayanan, G., Walker, C.~K., \& Buckley, H.~D.\ 1998, \apj, 496, 292 		
\bibitem[Narayanan \& Walker(1998)]{NarayananWalker1998} Narayanan, G., \& Walker, C.~K.\ 1998, \apj, 508, 780 
\bibitem[Pagani et al.(2009)]{Pagani2009} Pagani, L., Daniel, F., \& Dubernet, M.-L.\ 2009, \aap, 494, 719	
\bibitem[Rathborne et al.(2010)]{Rathborne2010}Rathborne, J.~M., Jackson, J.~M., Chambers, E.~T., et al.\ 2010, \apj, 715, 310
\bibitem[Rathborne et al.(2016)]{Rathborne2016}Rathborne, J.~M., Whitaker, J.~S., Jackson, J.~M., et al.\ 2016, \pasa, 33, e030	
\bibitem[Rudolph et al.(1990)]{Rudolph1990} Rudolph, A., Welch, W.~J., Palmer, P., \& Dubrulle, B.\ 1990, \apj, 363, 528 
\bibitem[Sanhueza et al.(2012)]{Sanhueza2012} Sanhueza, P., Jackson, J.~M., Foster, J.~B., et al.\ 2012, \apj, 756, 60 
\bibitem[Schmid-Burgk et al.(2004)]{Schmid-Burgk2004} Schmid-Burgk, J., Muders, D., M{\"u}ller, H.~S.~P., \& Brupbacher-Gatehouse, B.\ 2004, \aap, 419, 949
\bibitem[Schuller et al.(2009)]{Schuller2009} Schuller, F., Menten, K.~M., Contreras, Y., et al.\ 2009, \aap, 504, 415		
\bibitem[Shu(1977)]{Shu1977} Shu, F.~H.\ 1977, \apj, 214, 488		
\bibitem[Simpson \& Mayer-Hasselwander(1986)]{Simpson1986} Simpson, G., \& Mayer-Hasselwander, H.\ 1986, \aap, 162, 340
\bibitem[Smith et al.(2012)]{Smith2012} Smith, R.~J., Shetty, R., Stutz, A.~M., \& Klessen, R.~S.\ 2012, \apj, 750, 64 		
\bibitem[Smith et al.(2013)]{Smith2013} Smith, R.~J., Shetty, R., Beuther, H., Klessen, R.~S., \& Bonnell, I.~A.\ 2013, \apj, 771, 24
\bibitem[Snell \& Loren(1977)]{SnellLoren1977} Snell, R.~L., \& Loren, R.~B.\ 1977, \apj, 211, 122 
\bibitem[Stark(1981)]{Stark1981} Stark, A.~A.\ 1981, \apj, 245, 99 	
\bibitem[Traficante et al.(2018)]{Traficante2018} Traficante, A., Fuller, G.~A., Smith, R.~J., et al.\ 2018, \mnras, 473, 4975 		
\bibitem[Velusamy et al.(2008)]{Velusamy2008} Velusamy, T., Peng, R., Li, D., Goldsmith, P.~F., \& Langer, W.~D.\ 2008, \apjl, 688, L87 		
\bibitem[Walker et al.(1986)]{Walker1986} Walker, C.~K., Lada, C.~J., Young, E.~T., Maloney, P.~R., \& Wilking, B.~A.\ 1986, \apjl, 309, L47		
\bibitem[Welch et al.(1987)]{Welch1987} Welch, W.~J., Dreher, J.~W., Jackson, J.~M., Terebey, S., \& Vogel, S.~N.\ 1987, Science, 238, 1550		
\bibitem[Wyrowski et al.(2016)]{Wyrowski2016} Wyrowski, F., G{\"u}sten, R., Menten, K.~M., et al.\ 2016, \aap, 585, A149 		
\bibitem[Zhang et al.(1998)]{Zhang1988} Zhang, Q., Ho, P.~T.~P., \& Ohashi, N.\ 1998, \apj, 494, 636 		
\bibitem[Zhou(1992)]{Zhou1992} Zhou, S.\ 1992, \apj, 394, 204 		
\bibitem[Zhou et al.(1993)]{Zhou1993} Zhou, S., Evans, N.~J., II, Koempe, C., \& Walmsley, C.~M.\ 1993, \apj, 404, 232		
\bibitem[Zinnecker \& Yorke(2007)]{ZinneckerYorke2007} Zinnecker, H., \& Yorke, H.~W.\ 2007, \araa, 45, 481 

\end{thebibliography}
\end{document}